# Radiation pressure and the linear momentum of the electromagnetic field

**Masud Mansuripur**

*Optical Sciences Center, The University of Arizona, Tucson, Arizona 85721*
*masud@u.arizona.edu*

**Abstract:** We derive the force of the electromagnetic radiation on material objects by a direct application of the Lorentz law of classical electrodynamics. The derivation is straightforward in the case of solid metals and solid dielectrics, where the mass density and the optical constants of the media are assumed to remain unchanged under internal and external pressures, and where material flow and deformation can be ignored. For metallic mirrors, we separate the contribution to the radiation pressure of the electrical charge density from that of the current density of the conduction electrons. In the case of dielectric media, we examine the forces experienced by bound charges and currents, and determine the contribution of each to the radiation pressure. These analyses reveal the existence of a lateral radiation pressure inside the dielectric media, one that is exerted at and around the edges of a finite-diameter light beam. The lateral pressure turns out to be compressive for *s*-polarized light and expansive for *p*-polarized light. Along the way, we derive an expression for the momentum density of the light field inside dielectric media, one that has equal contributions from the traditional Minkowski and Abraham forms. This new expression for the momentum density, which contains both electromagnetic and mechanical terms, is used to explain the behavior of light pulses and individual photons upon entering and exiting a dielectric slab. In all the cases considered, the net forces and torques experienced by material bodies are consistent with the relevant conservation laws. Our method of calculating the radiation pressure can be used in conjunction with numerical simulations to yield the distribution of fields and forces in diverse systems of practical interest.

## 1. Introduction

It is well-known that the electromagnetic radiation carries both energy and momentum, that the energy flux is given by the Poynting vector $S$ of the classical electrodynamics, and that, in free-space, the momentum density $p$ (i.e., momentum per unit volume) is given by $p = S/c^2$, where $c$ is the speed of light in vacuum. What has been a matter of controversy for quite some time now is the proper form for the momentum of the electromagnetic waves in dielectric media. The question is whether the momentum density in a material medium has the form $p = D \times B$, due to Minkowski [1,2], or $p = E \times H/c^2$, due to Abraham [3,4]. J. P. Gordon [5] attributes the following comment to E. I. Blount: "The argument has not, it is true, been carried out at high volume, but the list of disputants is very distinguished." For a historical perspective on the subject and a summary of the relevant experimental results see R. Loudon [6,7] and J. P. Gordon [5].

Traditionally, the electromagnetic stress tensor has been used to derive the mechanical force exerted by the radiation field on ponderable media [8,9]. This approach, while having the advantage of generality, tends to obscure behind complicated mathematics the physical origin of the forces. It is possible, however, to calculate the force of the electromagnetic radiation on various media by direct invocation of the Lorentz law of force. The derivation is especially straightforward in the case of solid metals and solid dielectrics, where the mass density and the optical constants of the media may be assumed to remain constant under internal and external pressures, and where material flow and deformation can be ignored. Loudon [7] has emphasized "the simplicity and safety of calculations based on the Lorentz force and the dangers of calculations based on derived expressions involving elements of the Maxwell stress tensor, whose contributions may vanish in some situations but not in others."

In this paper we use the Lorentz law to derive the force of electromagnetic radiation on isotropic solid media in several simple situations. In the case of metallic mirrors, we separate, following Planck [10], the contribution to the radiation pressure of the electrical charge density from that of the current density (both due to conduction electrons). In the case of dielectric media, we examine the force experienced by bound charges and currents, and determine the contribution of each to the radiation pressure. Along the way, we derive a new expression for the momentum density of the light field inside dielectric media, one that has equal contributions from the aforementioned Minkowski and Abraham forms. This new expression for the momentum density, which contains both mechanical and electromagnetic terms, is subsequently used to elucidate the behavior of individual photons upon entering and exiting a dielectric slab. With the exception of the semi-quantitative results of Section 14, all the results obtained in this paper are exact, in the sense that no approximations or simplifications have been introduced; all derivations are based directly on the Lorentz law of force in conjunction with Maxwell's equations, using the standard constitutive relations for homogeneous, isotropic, linear, non-magnetic, and non-dispersive media.

The organization of the paper is as follows. In Section 2 we describe the notation and define the various parameters used throughout the paper. Section 3 considers the reflection of a plane electromagnetic wave from a perfect conductor, and shows that both the Lorentz law of force and the momentum of the electromagnetic field in free-space can consistently account for the radiation pressure exerted on the mirror surface. In Section 4 we use the Lorentz law to determine the radiation pressure on the surface of a semi-infinite dielectric medium at normal incidence. Here we derive an expression for the momentum of the field inside the dielectrics. The results of Section 4 are then extended to cover the case of oblique incidence on a semi-infinite dielectric, first with *s*-polarized light in Section 5, then with *p*-polarized light in Section 6. These analyses lead to the discovery of a lateral radiation pressure inside the dielectric medium, exerted at and around the edges of a finite-diameter plane wave. This lateral pressure, while having the same magnitude in both cases, turns out to be compressive for *s*-light and expansive for *p*-light. To the author's best knowledge, the expansive lateral force on the dielectric host of *p*-polarized light has not been discussed in the existing literature, making it a novel prediction that requires experimental verification.

In Section 7 we examine the torque experienced by a dielectric slab, illuminated at the Brewster's angle by a *p*-polarized plane wave. The torque is calculated directly from the Lorentz law applied to the induced (bound) charges at the surfaces of the slab, then shown to be consistent with the change in the angular momentum of the incident light. The case of an anti-reflection coated, semi-infinite dielectric medium is taken up in Section 8, where the increase in the momentum of the incident beam upon transmission into the dielectric medium is shown to result in a net force on the anti-reflection coating layer that tends to peel the layer away from its substrate; another new result that requires experimental verification. In Section 9 we analyze the case of a dielectric slab of finite thickness, and show that optical interference within the slab is responsible for the (longitudinal) stress induced by the electromagnetic radiation.

For a different perspective on the lateral pressure at the edges of a finite-diameter beam in a dielectric, Section 10 is devoted to an analysis of the one-dimensional Gaussian beam inside a dielectric medium. Depending on the direction of the *E*-field, we show that the lateral pressure on the medium can be compressive or expansive, and that the magnitude and direction of this radiation force are in complete accord with the results of Sections 5 and 6. The generality of this lateral pressure (and the dependence of its direction on the state of polarization) are brought to the fore in Section 11, where the simple fringes produced by the interference between two plane waves are shown to exhibit the same phenomena.

In Section 12 we extend our method of calculation of force and momentum to (finite-duration) light pulses, where the leading and trailing edges of the pulse are shown to play an important role in exchanging the electromagnetic momentum of the light with the mechanical momentum of the medium of propagation. The physical basis for the designation (in Section 4) of a fraction of the photon's momentum as "mechanical" is clarified in Section 12.

The classical experiments pertaining to metallic mirrors immersed in liquid dielectrics [11] are discussed in Section 13, where they are shown to be in complete agreement with our theoretical calculations. It is well known, e.g., from optical tweezers experiments [12-14], that a focused laser beam tends to attract small dielectric beads toward the center of the focused spot. At first glance, this observation might seem at odds with the presence of an expansive lateral pressure inside the dielectric medium of the bead. To resolve this apparent discrepancy, Section 14 offers a semi-quantitative analysis of a simplified model of the optical tweezers experiment. Final remarks and a summary of our important results appear in Section 15.

## 2. Notation and basic definitions

The MKSA system of units is used throughout the paper. Time harmonic fields are written as $\underline{E}(x, y, z, t) = E(x, y, z)\exp(-i\omega t)$, where $\omega = 2\pi f$ is the angular frequency. For brevity, we omit the explicit dependence of the fields on $x, y, z, t$. To specify their magnitude and phase, complex amplitudes such as $E$ are expressed as $|E|\exp(i\phi_E)$. Time-averaged products of two fields, say, $Real(\underline{A}) = |A|\cos(\omega t - \phi_A)$ and $Real(\underline{B}) = |B|\cos(\omega t - \phi_B)$, given by $\tfrac{1}{2}|AB|\cos(\phi_A - \phi_B)$, may also be written $\tfrac{1}{2}Real(AB^*)$.

To compute the force exerted by the electromagnetic field on a given medium, we use Maxwell's equations to determine the distributions of the $E$- and $H$-fields both inside and outside the medium. We then apply the Lorentz law $\boldsymbol{F} = q(\boldsymbol{E} + \boldsymbol{V} \times \boldsymbol{B})$, which gives the electromagnetic force on a particle of charge $q$ and velocity $\boldsymbol{V}$. The magnetic induction $\boldsymbol{B}$ is assumed to be related to the $H$-field via $\boldsymbol{B} = \mu_o \boldsymbol{H}$, where $\mu_o = 4\pi \times 10^{-7}$ henrys/meter is the permeability of free space; in other words, any magnetic moments that might exist in the medium and their interactions with the radiation field at optical frequencies are ignored. Typically, there are no free charges in the system, so $\nabla \cdot \boldsymbol{D} = 0$, where $\boldsymbol{D} = \varepsilon_o \boldsymbol{E} + \boldsymbol{P}$ is the electric displacement vector, $\varepsilon_o = 8.82 \times 10^{-12}$ farad/meter is the permittivity of free space, and $\boldsymbol{P}$ is the local polarization density within the medium. In linear media, $\boldsymbol{D} = \varepsilon_o \varepsilon \boldsymbol{E}$, where $\varepsilon$ is the medium's relative permittivity; hence, $\boldsymbol{P} = \varepsilon_o(\varepsilon - 1)\boldsymbol{E}$. We ignore the frequency-dependence of $\varepsilon$ throughout the paper and treat the media as non-dispersive.

When $\nabla \cdot \boldsymbol{D} = 0$, the density of bound charges $\rho_b = -\nabla \cdot \boldsymbol{P}$ may be expressed as $\rho_b = \varepsilon_o \nabla \cdot \boldsymbol{E}$. Inside a homogeneous and isotropic medium, $\boldsymbol{E}$ being proportional to $\boldsymbol{D}$ and $\nabla \cdot \boldsymbol{D} = 0$ imply that $\rho_b = 0$; no bound charges, therefore, can exist inside such media. However, at the interface between two different media, the component of $\boldsymbol{D}$ perpendicular to the interface, $\boldsymbol{D}_\perp$, must be continuous. The implication is that $\boldsymbol{E}_\perp$ is discontinuous and, therefore, bound charges can exist at such interfaces; the interfacial bound charges will thus have an areal density $\sigma = \varepsilon_o(E_{2\perp} - E_{1\perp})$. Under the influence of the local $E$-field, these charges give rise to an electric Lorentz force $\boldsymbol{F} = \frac{1}{2} Real(\sigma \boldsymbol{E}^*)$, where $\boldsymbol{F}$ is the force per unit area of the interface. Since the tangential $E$-field, $\boldsymbol{E}_\parallel$, is generally continuous across the interface, there is no ambiguity as to which field must be used in conjunction with the Lorentz law. As for the perpendicular component, the average $\boldsymbol{E}$ across the boundary, $\frac{1}{2}(\boldsymbol{E}_{1\perp} + \boldsymbol{E}_{2\perp})$, must be used in calculating the interfacial force. (The use of the average $\boldsymbol{E}_\perp$ in this context is not a matter of choice; it is the only way to get the calculated force at the boundary to agree with the time rate of change of the momentum that passes through the interface. From a physical standpoint, the interfacial charges produce a local $\boldsymbol{E}_\perp$ that has the same magnitude but opposite directions on the two sides of the interface. It is this locally-generated $\boldsymbol{E}_\perp$ that is responsible for the $E$-field's discontinuity. Averaging $\boldsymbol{E}_\perp$ across the interface eliminates the local field, as it should, since the charge cannot exert a force on itself.)

Since $\nabla \cdot \boldsymbol{B} = 0$ and $\boldsymbol{B} = \mu_o \boldsymbol{H}$, the perpendicular $H$-field, $\boldsymbol{H}_\perp$, at the interface between adjacent media must remain continuous. The tangential $H$-field at such interfaces, however, may be discontinuous. This, in accordance with Maxwell's equation $\nabla \times \underline{\boldsymbol{H}} = \underline{\boldsymbol{J}} + \partial \underline{\boldsymbol{D}}/\partial t$, gives rise to an interfacial current density $\boldsymbol{J}_s = \boldsymbol{H}_{2\parallel} - \boldsymbol{H}_{1\parallel}$. Such currents can exist on the surfaces of good conductors, where $\boldsymbol{E}_\parallel$ is negligible, yet the high conductance of the medium permits the flow of the surface current. Elsewhere, the only source of electrical currents are bound charges, with the bound current density being $\underline{\boldsymbol{J}}_b = \partial \underline{\boldsymbol{P}}/\partial t = \varepsilon_o(\varepsilon - 1)\partial \underline{\boldsymbol{E}}/\partial t$. Assuming time-harmonic fields with the time-dependence factor $\exp(-i\omega t)$, we can write $\boldsymbol{J}_b = -i\omega \varepsilon_o(\varepsilon - 1)\boldsymbol{E}$. The $H$-field of the electromagnetic wave then exerts a force on the bound current according to the Lorentz law, namely, $\boldsymbol{F} = \frac{1}{2} Real(\boldsymbol{J}_b \times \boldsymbol{B}^*)$, where $\boldsymbol{F}$ is force per unit volume.

**Note**: For time-harmonic fields, the contribution of conduction electrons to current density may be combined with that of bound electrons. Since $\boldsymbol{J}_c = \sigma_c \boldsymbol{E}$, where $\sigma_c$ is the conductivity of the medium, the net current density $\boldsymbol{J}_c + \boldsymbol{J}_b$ may be attributed to an effective dielectric constant $\varepsilon + i(\sigma_c / \varepsilon_o \omega)$. In general, since $\varepsilon$ is complex-valued, there is no need to distinguish conduction electrons from bound electrons, and $\varepsilon$ may be treated as an effective dielectric constant that contains both contributions. An exception will be made in Section 3 in the case of perfect conductors, where $\sigma_c \to \infty$. Here both $E$- and $H$-fields inside the medium tend to zero, and the contributions of bound charges/currents become negligible. The effect of the radiation in this case will be the creation of a surface current density $\boldsymbol{J}_s$ and a surface charge density $\sigma$, both of which may be attributed in their entirety to the conduction electrons. The conservation of charge then requires that $\nabla \cdot \boldsymbol{J}_s + \partial \underline{\sigma}/\partial t = 0$.

## 3. Reflection of plane wave from a perfect conductor

The material in this section is not new, but the line of reasoning and the methodology will be needed to build the necessary arguments in subsequent sections. The case of normal incidence on perfect conductors is well-known, but the two cases of oblique incidence, originally published in [10], have all but vanished from modern textbooks.

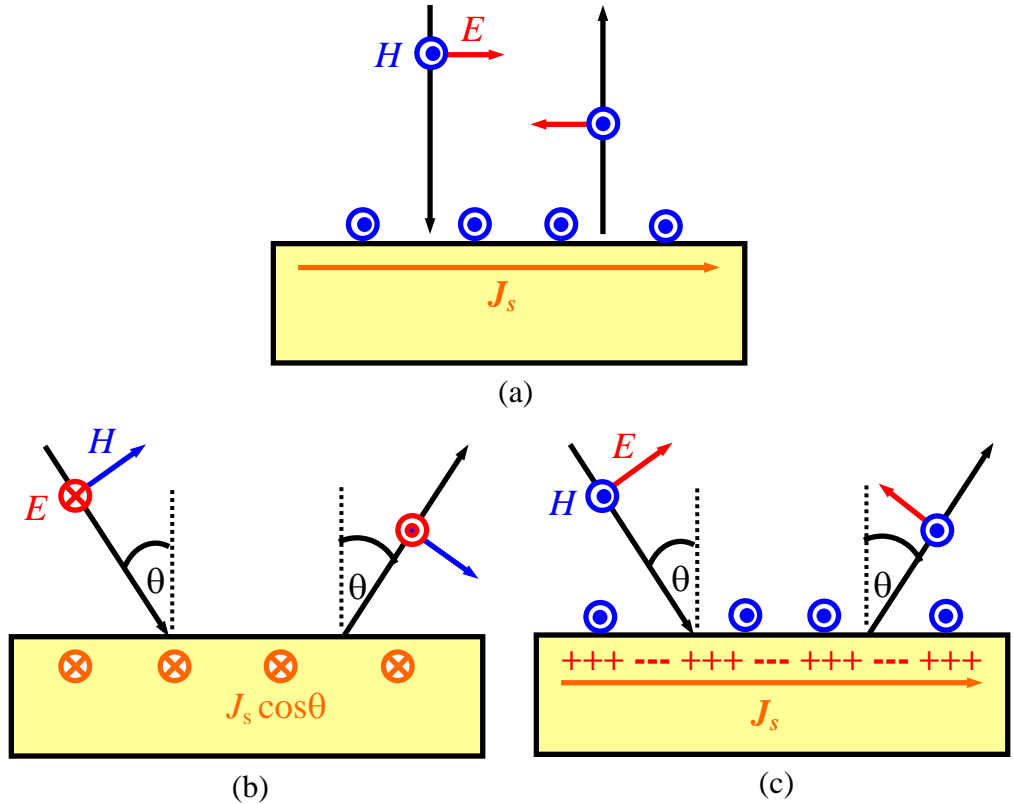

Fig. 1. A linearly-polarized plane wave is reflected from a perfectly conducting mirror. Whereas the parallel component of the *E*-field at the mirror surface is zero, the parallel component of the *H*-field is at its maximum. The surface current $\boldsymbol{J}_s$ is equal in magnitude and perpendicular in direction to the magnetic field at the surface. (a) Normal incidence. (b) Oblique incidence with *s*-polarization. (c) Oblique incidence with *p*-polarization.

In Fig.1(a) a plane wave of wavelength $\lambda_o$, having *E*-field amplitude $E_o$ (units = V/m) and *H*-field amplitude $H_o = E_o/Z_o$ (units = A/m), where $Z_o = \sqrt{\mu_o/\varepsilon_o} \sim 377\Omega$ is the free-space impedance, is incident on a perfect conductor. The Poynting vector is $\boldsymbol{S} = \frac{1}{2}Real(\boldsymbol{E} \times \boldsymbol{H}^*)$, and the momentum density (per unit volume) is $\boldsymbol{p} = \boldsymbol{S}/c^2$ (vacuum speed of light $c = 1/\sqrt{\mu_o \varepsilon_o}$). In unit time, the incoming momentum over a unit area of the reflector is that contained in a column of base $A = 1.0 \text{ m}^2$ and height $c$. The same momentum returns to the source after being reflected from the mirror, so the net rate of change of the field momentum over a unit area $d\boldsymbol{p}/dt = 2\boldsymbol{S}/c$, which is equal to the force per unit area, $\boldsymbol{F}$, exerted on the reflector. The force density of the light on a perfect reflector in free-space is thus given by

$$\bm{F} = \varepsilon_o E_o^2. \qquad (1)$$

The same expression may be derived by considering the Lorentz law $\bm{F} = q\,(\bm{E} + \bm{V} \times \bm{B})$ in conjunction with the surface current density $\bm{J}_s$ and the magnetic field $\bm{H}$ at the surface of the conductor. Here there are neither free nor (unbalanced) bound charges, and the motion of the conduction electrons constitutes a surface current density $\bm{J}_s = q\bm{V}$. For time-harmonic fields, the force per unit area may thus be written

$$\bm{F} = \tfrac{1}{2} Real(\bm{J_s} \times \bm{B}^*). \qquad (2)$$

In the case of a perfect conductor the magnitude of the surface current is equal to the magnetic field at the mirror surface, namely, $J_s = 2H_o$ (because $\nabla \times \underline{\bm{H}} = \underline{\bm{J}} + \partial\underline{\bm{D}}/\partial t$; the factor of 2 arises from the interference between the incident and reflected beams where the two $H$-fields, being in-phase at the mirror surface, add up.) Since $\bm{B} = \mu_o\bm{H}$, one might conclude that $F_z = 2\mu_o H_o^2$. The factor of 2 in this formula, however, is incorrect because the magnetic field on the film's surface, $2H_o$, is assumed to exert a force on the *entire* $J_s$. The problem is that the field is $2H_o$ at the top of the mirror and zero just under the surface, say, below the skin-depth. (Here we are using a limiting argument in which a good conductor, having a finite skin-depth, approaches an ideal conductor in the limit of zero skin-depth.) Therefore, the average $H$-field through the "skin-depth" must be used in calculating the force, and this average is $H_o$ not $2H_o$. The force per unit area thus calculated is $F_z = \mu_o H_o^2 = \varepsilon_o E_o^2$, which is identical to the time rate of change of momentum of the incident beam derived in Eq. (1). With 1.0 W/mm$^2$ of incident optical power, for example, the radiation pressure on the mirror will be 6.67 nN/mm$^2$.

Next, suppose the beam arrives on the mirror at an oblique angle θ, as in Fig. 1(b); here the beam is assumed to be *s*-polarized. Compared to normal incidence, the component of the magnetic field $H$ on the surface is now multiplied by cosθ, which requires the surface current density $J_s$ to be multiplied by the same factor (remember that $J_s$ is equal to the magnetic field at the surface). The component of force density along the *z*-axis, $F_z$, is thus seen to have been reduced by a factor of cos$^2$θ. This result is consistent with the alternative derivation based on the time rate of change of the field's momentum in the *z*-direction, d$p_z$/d$t$, which is multiplied by cosθ in the case of oblique incidence. Since the beam has a finite diameter, its foot-print on the mirror is greater than that in the case of normal incidence by 1/cosθ. Thus the force density $F_z$, obtained by normalizing d$p_z$/d$t$ by the beam's foot-print, is seen once again to be reduced by a factor of cos$^2$θ.

Figure 1(c) shows a *p*-polarized beam at oblique incidence on a mirror. The magnetic field component at the surface is $2H_o$, which means that the surface current $J_s$ must also have the same magnitude as in normal incidence. We conclude that the force density on the mirror must be the same as that at normal incidence, namely, $F_z = \varepsilon_o E_o^2$. The time rate of change of momentum in the *z*-direction, however, is similar to that in Fig. 1(b), which means that the force density of normal incidence must have been multiplied by cos$^2$θ in the case of oblique incidence. The two methods of calculating $F_z$ for *p*-light thus disagree by a factor of cos$^2$θ.

The discrepancy is resolved when one realizes that, in addition to the magnetic force, an electric force is acting on the mirror in the opposite direction (−*z*). This additional force pulls on the electric charges induced at the surface by $E_\perp$. Note that $E_\perp = 2E_o\sin\theta$ just above and $E_\perp = 0$ just below the surface. The discontinuity in $E_\perp$ gives the surface charge density as $\sigma = 2\varepsilon_o E_o\sin\theta$. The perpendicular *E*-field acting on these charges is the average of the fields just above and just below the surface, namely, $E_z^{(eff)} = E_o\sin\theta$. The electric force density is thus $F_z = \tfrac{1}{2} Real\,(\sigma E_z^{(eff)*}) = \varepsilon_o E_o^2\sin^2\theta$. The upward force on the charges thus reduces the downward force on the current, leading to a net $F_z = \varepsilon_o E_o^2(1 - \sin^2\theta)$, which is the same as that in the case of normal incidence multiplied by cos$^2$θ.

The charge density $\sigma = 2\varepsilon_o E_o \sin\theta \exp(i2\pi x \sin\theta/\lambda_o)$ in the above example is produced by the spatial variations of the current density $J_s = 2H_o\exp(i2\pi x \sin\theta/\lambda_o)$. Conservation of charge requires $\nabla \cdot \boldsymbol{J} + \partial\rho/\partial t = 0$, which, for time-harmonic fields, reduces to $\partial J_s/\partial x - i\omega\sigma = 0$. Considering that $H_o = E_o/Z_o$ and $\omega = 2\pi c/\lambda_o$, it is readily seen that the above $J_s$ and $\sigma$ satisfy the required conservation law.

**Note**: The separate contributions of charge and current to the radiation pressure discussed in this section were originally discussed by Max Planck in his 1914 book, *The Theory of Heat Radiation* [10]. Our brief reconstruction of his arguments here is intended to facilitate the following discussion of electromagnetic force and momentum in dielectric media.

### 4. Semi-infinite dielectric

This section presents the core of the argument that leads to a new expression for the momentum of light inside dielectric media. Loudon [6,7] has presented a similar argument in his quantum mechanical treatment of the problem. Although Loudon's final result comes close to ours, there are differences that can be traced to his neglect of the mechanism of photon entry from the free-space into the dielectric medium.

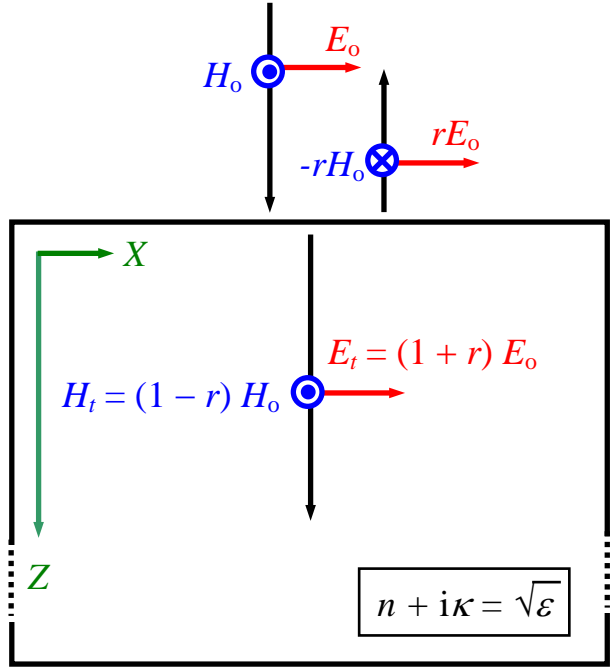

Fig. 2. A linearly-polarized plane wave is normally incident on the surface of a semi-infinite medium of complex dielectric constant $\varepsilon$. The Fresnel reflection coefficient at the surface is $r$. Shown are the *E*- and *H*-field magnitudes for the incident, reflected, and transmitted beams.

Figure 2 shows a linearly-polarized plane wave at normal incidence on the flat surface of a semi-infinite dielectric. The incident *E*- and *H*-fields have magnitudes $E_o$ and $H_o = E_o/Z_o$. Assuming a beam cross-sectional area of unity ($A = 1.0\text{m}^2$), the time rate of flow of momentum onto the surface is $\tfrac{1}{2}\varepsilon_o E_o^2$, of which a fraction $|r|^2$ is reflected back. The net rate of change of linear momentum, which must be equal to the force per unit area exerted on the surface, is thus $F_z = \tfrac{1}{2}\varepsilon_o(1 + |r|^2)E_o^2$. We assume that the medium's dielectric constant $\varepsilon$ is not purely real, but has a small imaginary part. The complex refractive index of the material is $n + i\kappa = \sqrt{\varepsilon}$, and the reflection coefficient is $r = (1 - \sqrt{\varepsilon})/(1 + \sqrt{\varepsilon})$.

Inside the dielectric, the $E$-field is $E(z) = E_t \exp(i2\pi\sqrt{\varepsilon}\,z/\lambda_o)$, where $E_t = (1 + r)E_o$, the $H$-field is $H(z) = \sqrt{\varepsilon}(E_t/Z_o)\exp(i2\pi\sqrt{\varepsilon}\,z/\lambda_0)$, the $D$-field is $D(z) = \varepsilon_o E(z) + P(z) = \varepsilon_o \varepsilon E(z)$, and the dipolar current density is $J(z) = -i\omega P(z) = -i\omega\varepsilon_o(\varepsilon - 1)E(z)$, where $\omega = 2\pi f = 2\pi c/\lambda_o$ is the optical frequency. The force per unit volume is thus given by

$$F_z = \tfrac{1}{2}\,Real\,(\boldsymbol{J}\times\boldsymbol{B}^*) = \tfrac{1}{2}(2\pi/\lambda_o)\,Real\,[-i\sqrt{\varepsilon^*}(\varepsilon-1)]\,\varepsilon_o|E_t|^2\exp(-4\pi\kappa z/\lambda_o)$$
$$= \tfrac{1}{2}(2\pi/\lambda_o)(n^2 + \kappa^2 + 1)\kappa\,\varepsilon_o|E_t|^2\exp(-4\pi\kappa z/\lambda_o). \tag{3}$$

The total force per unit surface area is obtained by integrating the above $F_z$ from $z = 0$ to $\infty$. The multiplicative coefficient $\kappa$ disappears after integration, and the force per unit area becomes $F_z = \tfrac{1}{4}(n^2 + \kappa^2 + 1)\varepsilon_o|E_t|^2$. Upon substitution for $E_t$ and $r$, this expression for $F_z$ turns out to be identical to that obtained earlier based on momentum considerations.

We now let $\kappa \to 0$ and write the radiation force per unit surface area of the dielectric as $F_z = \tfrac{1}{4}(n^2 + 1)\varepsilon_o|E_t|^2$. (A similar trick has been used by R. Loudon in his calculation of the photon momentum inside dielectrics [7].) Considering that $H_t = nE_t/Z_o$, one may also write $F_z = \tfrac{1}{4}\varepsilon_o|E_t|^2 + \tfrac{1}{4}\mu_o|H_t|^2$. This must be equal to the rate of the momentum entering the medium at $z = 0$. Since the speed of light in the medium is $c/n$, the momentum density (per unit volume) within the dielectric may be expressed as follows:

$$p_z = \tfrac{1}{4}(n^2 + 1)n\varepsilon_o|E_t|^2/c = \tfrac{1}{4}(\varepsilon + 1)\varepsilon_o|E_t B_t|. \tag{4}$$

Equation (4), the fundamental expression for the momentum density of plane waves in dielectrics, may also be written as $\boldsymbol{p} = \tfrac{1}{4}(\boldsymbol{D}\times\boldsymbol{B}) + \tfrac{1}{4}(\boldsymbol{E}\times\boldsymbol{H})/c^2$. Historically, there has been a dispute as to whether the proper form for the momentum density of light in dielectrics is Minkowski's $\tfrac{1}{2}\boldsymbol{D}\times\boldsymbol{B}$ or Abraham's $\tfrac{1}{2}\boldsymbol{E}\times\boldsymbol{H}/c^2$ [5]. The above discussion leads to the conclusion that neither form is appropriate; rather, it is the average of the two that yields the most plausible expression for $\boldsymbol{p}$. In the limit when $\varepsilon \to 1$, the two terms in the expression for $\boldsymbol{p}$ become identical, and the familiar form for the free-space, $\boldsymbol{p} = \boldsymbol{S}/c^2$, emerges.

Replacing $\boldsymbol{D}$ with $\varepsilon_o\boldsymbol{E} + \boldsymbol{P}$ and $\boldsymbol{B}$ with $\mu_o\boldsymbol{H}$, we obtain $\boldsymbol{p} = \tfrac{1}{4}(\boldsymbol{P}\times\boldsymbol{B}) + \tfrac{1}{2}(\boldsymbol{E}\times\boldsymbol{H})/c^2$, which shows the separate contributions to a plane-wave's momentum density by the medium and by the radiation field. The mechanical momentum of the medium, $\tfrac{1}{4}\boldsymbol{P}\times\boldsymbol{B}$, arises from the interaction between the induced polarization density $\boldsymbol{P}$ and the light's $\boldsymbol{B}$-field. The contribution of the radiation field, $\tfrac{1}{2}\boldsymbol{E}\times\boldsymbol{H}/c^2$, has the same form, $\boldsymbol{S}/c^2$, as the momentum density of electromagnetic radiation in free space. Since $\boldsymbol{P} = \varepsilon_o(\varepsilon - 1)\boldsymbol{E}$, the mechanical momentum density may be written as $\tfrac{1}{4}\boldsymbol{P}\times\boldsymbol{B} = \tfrac{1}{2}(\varepsilon - 1)\boldsymbol{S}/c^2$. For a dilute medium having refractive index $n \approx 1$, the coefficient of $\boldsymbol{S}/c^2$ in the above formula reduces to $\tfrac{1}{2}(\varepsilon - 1) \approx n - 1$, which leads to the expression $(n-1)\boldsymbol{S}/c^2$ derived in [5] for the mechanical momentum of dilute gases. The physical basis for the separation of the momentum density into electromagnetic and mechanical contributions will be further elaborated in Section 12.

**Note**: In a recent paper [15], Obukhov and Hehl argue, as we do here, that the correct interpretation of the electromagnetic momentum in dielectric media must be based on the standard form of the Lorentz force, taking into account both free and bound charges and currents. In their discussion of the case of normal incidence from vacuum onto a semi-infinite dielectric, however, they neglect to account for the mechanical momentum imparted to the dielectric medium. As a result, they find only the electromagnetic part of the momentum density; their Eq. (27) is in fact identical to $\tfrac{1}{2}\boldsymbol{E}\times\boldsymbol{H}/c^2$, where $\boldsymbol{E}$ and $\boldsymbol{H}$ are evaluated inside the dielectric. In contrast, our approach in the present section, which involves the introduction of a small (but non-zero) $\kappa$, followed by an integration of the feeble magnetic Lorentz force over the infinite thickness of the dielectric, ensures that the mechanical momentum of the medium is properly taken into consideration. This yields the term $\tfrac{1}{4}(\boldsymbol{P}\times\boldsymbol{B})$ in our last expression for the momentum density, which is missing from Obukhov and Hehl's Eq. (27).

## 5. Oblique incidence with *s*-polarized light

To the best of our knowledge, the momentum of light at oblique incidence has not been discussed previously. This is an extremely important case, since it reveals the existence of a lateral radiation pressure at the edges of the beam within a dielectric medium; consistency with the results of Section 4 simply demands the existence of such lateral pressures. We analyze the case of *s*-polarization here, leaving a discussion of *p*-polarized light at oblique incidence for the next section. The two cases turn out to be fundamentally different, although both retain the expression for momentum density derived in the case of normal incidence.

Figure 3 shows the case of oblique incidence with *s*-polarized light at the interface between the free-space and a dielectric medium. Again, we assume that the dielectric constant $\varepsilon$ is complex, allowing it to approach a real number only after calculating the total force by integrating through the thickness of the medium. Inside the medium, the *E*- and *H*-field distributions are

$$E_{ty}(x, z) = (1 + r_s)E_o \exp[i2\pi(x\sin\theta + z\sqrt{\varepsilon - \sin^2\theta})/\lambda_o] \tag{5a}$$

$$H_{tx}(x, z) = \sqrt{\varepsilon - \sin^2\theta}\, E_{ty}(x, z)/Z_o \tag{5b}$$

$$H_{tz}(x, z) = -\sin\theta\, E_{ty}(x, z)/Z_o \tag{5c}$$

Here $r_s = (\cos\theta - \sqrt{\varepsilon - \sin^2\theta})/(\cos\theta + \sqrt{\varepsilon - \sin^2\theta})$ is the Fresnel reflection coefficient for *s*-light. Since there are no free charges inside the medium (nor on its surface), the only relevant force here is the magnetic Lorentz force on the dipolar current density $J_y(x, z) = -i\omega\varepsilon_o(\varepsilon - 1)E_{ty}(x, z)$. Following the same procedure as before, we find the net force components along the *x*- and *z*-axes to be

$$F_x = \tfrac{1}{2}\varepsilon_o \sin\theta\, Real(\sqrt{\varepsilon - \sin^2\theta})\,|1 + r_s|^2 E_o^2 \tag{6a}$$

$$F_z = \tfrac{1}{4}\varepsilon_o(\cos^2\theta + |\varepsilon - \sin^2\theta|)\,|1 + r_s|^2 E_o^2 \tag{6b}$$

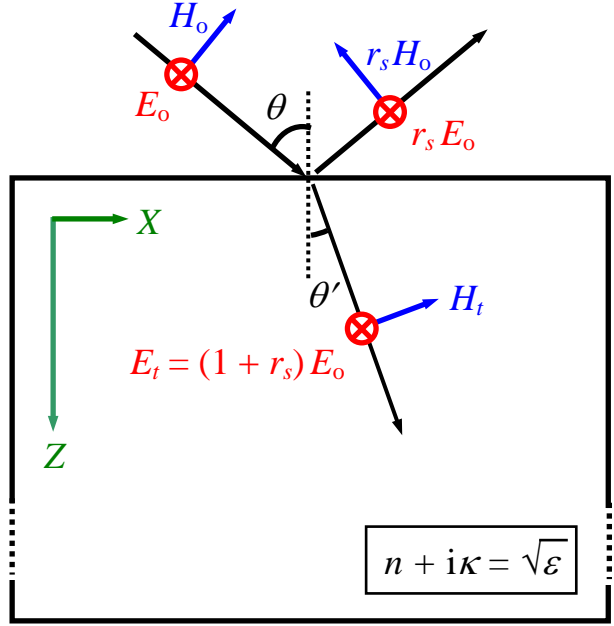

Fig. 3. Obliquely incident *s*-polarized plane wave arrives at the surface of a semi-infinite medium of (complex) dielectric constant $\varepsilon$. The Fresnel reflection coefficient is denoted by $r_s$.

The above results are valid for all values of $\varepsilon$, whether real or complex. Considering that the incident beam's cross-sectional area must be $A = \cos\theta$ to produce a unit area footprint at the interface, the time rate of change of the incident beam's momentum upon reflection from the surface gives rise to $F_x = \tfrac{1}{2}\varepsilon_o \sin\theta \cos\theta \,(1 - |r_s|^2)E_o^2$ and $F_z = \tfrac{1}{2}\varepsilon_o \cos^2\theta \,(1 + |r_s|^2)E_o^2$. These can be readily shown to agree with Eq. (6), which has been obtained by direct integration of the force density through the thickness of the medium.

Next we allow $\varepsilon$ to be real, and set the refractive index $n = \sqrt{\varepsilon}$. Since $|E_t|=|1+r_s|^2 E_o^2$, $\sin\theta = n\sin\theta'$, and $\sqrt{\varepsilon - \sin^2\theta} = n\cos\theta'$, we may write Eq. (6) as follows:

$$F_x = \tfrac{1}{2}\varepsilon_o \varepsilon \sin\theta' \cos\theta' \,|E_t|^2, \tag{7a}$$

$$F_z = \tfrac{1}{4}\varepsilon_o (1 - \varepsilon \sin^2\theta' + \varepsilon \cos^2\theta')\,|E_t|^2. \tag{7b}$$

Note that the cross-sectional area $A$ of the transmitted beam is not unity, but $\cos\theta'$, which means that the above expressions for $F_x$ and $F_z$ must be divided by $\cos\theta'$ if force per unit area is desired. The direction of the force $\boldsymbol{F} = F_x \boldsymbol{x} + F_z \boldsymbol{z}$ is *not* the same as the propagation direction (i.e., at angle $\theta'$ to the surface normal); the reason for this will become clear shortly.

From Section 4 we know that, inside the dielectric, the force per unit area along the propagation direction must be $\boldsymbol{F} = \tfrac{1}{4}\varepsilon_o(\varepsilon + 1)|E_t|^2(\sin\theta'\boldsymbol{x} + \cos\theta'\boldsymbol{z})$. Multiplying this force by the beam's cross-sectional area $A = \cos\theta'$, then subtracting it from the previously calculated force in Eq. (7) yields the following residual force:

$$\Delta\boldsymbol{F} = \tfrac{1}{4}\varepsilon_o(\varepsilon - 1)\sin\theta'(\cos\theta'\boldsymbol{x} - \sin\theta'\boldsymbol{z})\,|E_t|^2. \tag{8}$$

The force in Eq. (8) is orthogonal to the beam's propagation direction, $\sin\theta'\boldsymbol{x} + \cos\theta'\boldsymbol{z}$. Moreover, the magnitude of $\Delta\boldsymbol{F}$ is proportional to $\sin\theta'$, the cross-sectional area of a segment from the edge of the beam just below the interface; see Fig. 4. The force per unit area of the beam's edge is thus $F^{(edge)} = \tfrac{1}{4}\varepsilon_0(\varepsilon - 1)|E_t|^2$. (For example, if the optical power density inside a glass medium having $n = 1.5$ happens to be 1.0 W/mm$^2$, the lateral pressure on each edge of the beam will be 1.39 nN/mm$^2$.) This force, which the light exerts on the dielectric at both edges of the beam, is orthogonal to the edge and compressive in this case of *s*-polarized light. (We will see in the next section that the force at the edges of a *p*-polarized beam has exactly the same magnitude but opposite direction, tending to expand the dielectric medium.) Once the component of the force acting on the beam's edge has been subtracted from Eq. (7), the remaining force turns out to be in the propagation direction and in full agreement with the results of the preceding section.

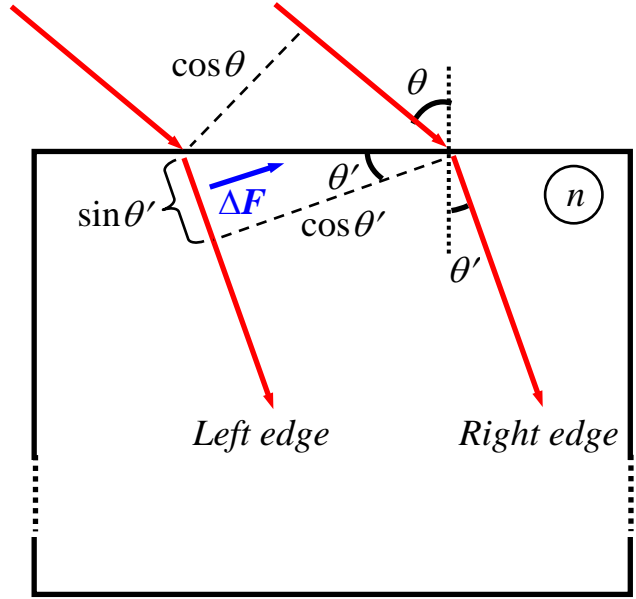

Fig. 4. Oblique incidence on a semi-infinite dielectric at angle $\theta$. The beam's footprint at the surface has unit area, while the incident and transmitted beams' cross-sectional areas are $\cos\theta$ and $\cos\theta'$, respectively. A segment from the beam's left edge (area proportional to $\sin\theta'$) exerts a force $\Delta\boldsymbol{F}$ on the dielectric; this force is not compensated by an equal and opposite force on the right-hand edge of the beam, as is the case elsewhere at the opposite edges of the beam. The compressive $\Delta\boldsymbol{F}$ shown here corresponds to an *s*-polarized beam. (For *p*-light $\Delta\boldsymbol{F}$ retains the same magnitude but reverses direction, so the edge force becomes expansive.)

## 6. Oblique incidence with *p*-polarized light

The case of *p*-polarized light depicted in Fig. 5 differs somewhat from the case of *s*-polarized light in that, in addition to the magnetic Lorentz force on its bulk, the medium also experiences an electric Lorentz force on the (bound) charges induced at its interface with the free space. Inside the medium the field distributions are

$$H_{ty}(x, z) = (1 - r_p) H_o \exp[i2\pi(x \sin\theta + z\sqrt{\varepsilon - \sin^2\theta})/\lambda_o], \qquad (9a)$$

$$E_{tx}(x, z) = (\sqrt{\varepsilon - \sin^2\theta}/\varepsilon) Z_o H_{ty}(x, z), \qquad (9b)$$

$$E_{tz}(x, z) = -(\sin\theta/\varepsilon) Z_o H_{ty}(x, z). \qquad (9c)$$

Here $r_p = (\sqrt{\varepsilon - \sin^2\theta} - \varepsilon\cos\theta)/(\sqrt{\varepsilon - \sin^2\theta} + \varepsilon\cos\theta)$ is the Fresnel reflection coefficient at the interface for *p*-light. Since there are no net bound charges inside the medium, the only relevant force in the bulk is the Lorentz force of the *H*-field on the dipolar current density $\boldsymbol{J}(x, z) = -i\omega\varepsilon_0(\varepsilon - 1)\boldsymbol{E}_t(x, z)$. Following the same procedure as before, we find the force components exerted on the bulk along the *x*- and *z*-axes as follows:

$$F_x^{(bulk)} = \tfrac{1}{2}\mu_o \sin\theta \, Real(\sqrt{\varepsilon - \sin^2\theta})\,|\varepsilon|^{-2}\,|1 - r_p|^2 H_o^2, \qquad (10a)$$

$$F_z^{(bulk)} = \tfrac{1}{4}\mu_o(|\varepsilon|^2 - \sin^2\theta + |\varepsilon - \sin^2\theta|)\,|\varepsilon|^{-2}\,|1 - r_p|^2 H_o^2. \qquad (10b)$$

Next we consider the density of bound charges at the interface with the free space. In the region just above the interface the *z*-component of $\boldsymbol{E}$ is $E_z(x, z= 0) = (1 - r_p)\sin\theta\, E_o \exp(i2\pi x \sin\theta/\lambda_o)$. Immediately below the interface, the continuity of $\boldsymbol{D}_\perp$ requires that the above $E_z$ be divided by $\varepsilon$. The surface charge density $\sigma$, being equal to the discontinuity in $\varepsilon_o E_z$, is thus given by

$$\sigma = \varepsilon_o(1 - 1/\varepsilon)(1 - r_p)\sin\theta\, E_o \exp(i2\pi x \sin\theta/\lambda_o). \qquad (11)$$

The force components felt by these charges are $F_x = \tfrac{1}{2} Real\,[\sigma(x)E_x^*(x, z=0)]$, where $E_x$ is continuous across the interface, and $F_z = \tfrac{1}{2} Real\,[\sigma(x)E_z^*(x, z = 0)]$, where $E_z$ is the average $E_\perp$ across the boundary, namely, $E_z = \tfrac{1}{2}(1 + 1/\varepsilon)(1 - r_p)\sin\theta\, E_o \exp(i2\pi x \sin\theta/\lambda_o)$. Thus

$$F_x^{(surface)} = \tfrac{1}{2}\mu_o \sin\theta\, Real\,[(\varepsilon^* - 1)\sqrt{\varepsilon - \sin^2\theta}\,]\,|\varepsilon|^{-2}\,|1 - r_p|^2 H_o^2 \qquad (12a)$$

$$F_z^{(surface)} = -\tfrac{1}{4}\mu_o \sin^2\theta\,(1 - |\varepsilon|^{-2})\,|1 - r_p|^2 H_o^2 \qquad (12b)$$

The total force, the sum of the bulk and surface forces of Eqs. (10) and (12), is given by

$$F_x^{(total)} = \tfrac{1}{2}\mu_o \sin\theta\, Real\,(\varepsilon^*\sqrt{\varepsilon - \sin^2\theta}\,)\,|\varepsilon|^{-2}\,|1 - r_p|^2 H_o^2 \qquad (13a)$$

$$F_z^{(total)} = \tfrac{1}{4}\mu_o(\,|\varepsilon|^2\cos^2\theta + |\varepsilon - \sin^2\theta|\,)|\varepsilon|^{-2}\,|1 - r_p|^2 H_o^2 \qquad (13b)$$

These results are valid for all values of $\varepsilon$, real or complex. Considering that the incident beam's cross-sectional area must be $A = \cos\theta$ to yield a unit-area footprint at the interface, the time rate of change of the incident beam's momentum upon reflection gives rise to $F_x = \tfrac{1}{2} \sin\theta \cos\theta (1 - |r_p|^2)\mu_o H_o^2$ and $F_z = \tfrac{1}{2}\cos^2\theta (1 + |r_p|^2)\mu_o H_o^2$. These can easily be shown to agree with Eqs. (13), obtained by a direct calculation of the force from the Lorentz law.

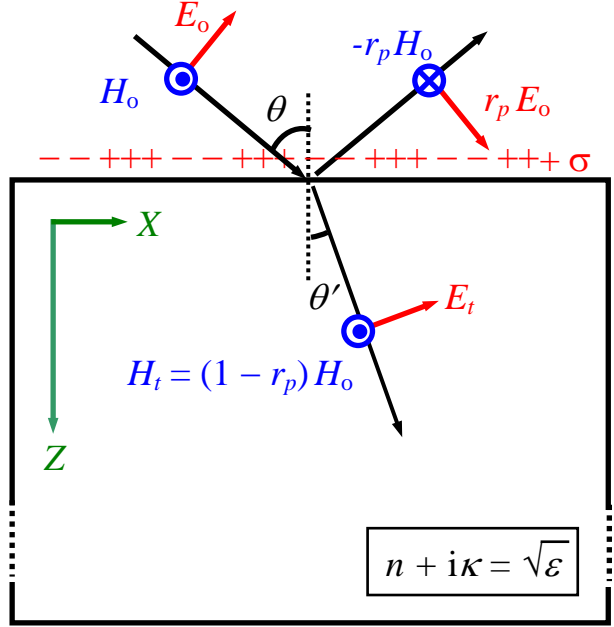

Fig. 5. A *p*-polarized plane wave is obliquely incident at the surface of a semi-infinite medium of (complex) dielectric constant $\varepsilon$. The Fresnel reflection coefficient is denoted by $r_p$.

It is of some interest to compare the forces of *s*- and *p*-beams on the bulk of the medium given by Eqs. (6) and (10). In Eq. (6) the *E*-field inside the medium is $E_t = (1 + r_s)E_o$, whereas in Eq. (10) the *H*-field in the medium is $H_t = (1 - r_p)H_o$. From Eqs. (5) and (9) we find

$$\sqrt{\varepsilon}\, E_{ty}/Z_o = \sqrt{H_{tx}^2 + H_{tz}^2} \tag{14a}$$

$$H_{ty} = \sqrt{\varepsilon E_{tx}^2 + \varepsilon E_{tz}^2}/Z_o. \tag{14b}$$

Thus, even when $|E_t|$ is the same in both cases, the force components turn out to be different. This is caused by the direction of the force on the beam's edges being different in the two cases, as will become clear below when we analyze the case of media with real-valued $\varepsilon$.

We now allow $\varepsilon$ to be real, and set the refractive index $n = \sqrt{\varepsilon}$. Since $|H_t|^2 = |1 - r_p|^2 H_o^2$, $H_t = nE_t/Z_o$, $\sin\theta = n\sin\theta'$, and $\sqrt{\varepsilon - \sin^2\theta} = n\cos\theta'$, Eq. (10) can be written

$$F_x^{(\text{bulk})} = \tfrac{1}{2}\varepsilon_o \sin\theta' \cos\theta' |E_t|^2, \tag{15a}$$

$$F_z^{(\text{bulk})} = \tfrac{1}{4}\varepsilon_o(\varepsilon - \sin^2\theta' + \cos^2\theta') |E_t|^2. \tag{15b}$$

Note that the cross-sectional area $A$ of the transmitted beam is not unity, but $\cos\theta'$. From the earlier discussions we know that, inside the dielectric, the force per unit cross-sectional area of the beam is $\boldsymbol{F} = \tfrac{1}{4}\varepsilon_o(\varepsilon + 1)|E_t|^2(\sin\theta'\boldsymbol{x} + \cos\theta'\boldsymbol{z})$, aligned with the propagation direction. Multiplying this force by the beam's cross-sectional area $A = \cos\theta'$, then subtracting it from the previously calculated force in Eq. (15), yields the following residual force:

$$\Delta \boldsymbol{F} = -\tfrac{1}{4}\varepsilon_o(\varepsilon - 1)\sin\theta'(\cos\theta'\boldsymbol{x} - \sin\theta'\boldsymbol{z})|E_t|^2. \tag{16}$$

The force in Eq. (16) is orthogonal to the beam's propagation direction, $\sin\theta'\boldsymbol{x} + \cos\theta'\boldsymbol{z}$. Moreover, its magnitude is proportional to $\sin\theta'$, the cross-sectional area of a segment from the edge of the beam just below the interface; see Fig. 4. The force per unit area of the beam's edge is thus $F^{(\text{edge})} = \tfrac{1}{4}\varepsilon_o(\varepsilon - 1)|E_t|^2$. This force, exerted on the dielectric at both edges of the beam, is orthogonal to the edge and expansive in this case of *p*-polarized light. Once the force component acting on the beam's edge has been subtracted from Eq. (15), the remaining force turns out to be in the propagation direction and in full agreement with the results of Section 4.

## 7. Dielectric slab illuminated at Brewster's angle

This problem has been studied by Barlow [16], although not at Brewster's angle. The simplification afforded by our choice of the incidence angle is the elimination of multiple reflections within the slab. Unlike Barlow, we focus our attention on the radiation pressure at the edges of the beam inside the dielectric, and obtain the same result as in Section 6 from a completely new and unexpected perspective.

A *p*-polarized plane wave is incident on a dielectric slab of thickness $d$ and refractive index $n$ at the Brewster's angle $\theta_B$ ($\tan\theta_B = n$); the refracted angle inside the slab is given by $\tan\theta'_B = 1/n$. Since the reflectivity at Brewster's angle is zero, the only beams in this system are the incident beam, the refracted beam inside the slab, and the transmitted beam; see Fig. 6(a). Inside the slab, the *H*-field is the same as that outside, as required by the continuity of $\boldsymbol{H}_\parallel$ at the interfaces. Similarly, the continuity of $\boldsymbol{E}_\parallel$ and $\boldsymbol{D}_\perp$ at the interfaces require that, inside the slab and just beneath the surface, $E_x = E_o\cos\theta_B$, and $E_z = (E_o/n^2)\sin\theta_B$, which means that the magnitude of $E$ inside the slab is $E_o/n$.

Inside the slab, the density of bound charges is zero, because $\nabla\cdot\boldsymbol{D} = \varepsilon_o\varepsilon\nabla\cdot\boldsymbol{E} = 0$. The bound currents have a density of $\underline{\boldsymbol{J}}_b = \partial\underline{\boldsymbol{P}}/\partial t = \varepsilon_o(\varepsilon - 1)\partial\underline{\boldsymbol{E}}/\partial t$. However, the force exerted by the magnetic field of the light on these currents is zero because of the 90° phase between $\underline{\boldsymbol{H}}$ and $\partial\underline{\boldsymbol{E}}/\partial t$. Thus the Lorentz force inside the volume of the slab is zero. The only force is due to the bound charges induced on the two surfaces of the plate, the density of which is obtained from the discontinuity in $\varepsilon_o E_\perp$ across the interface, namely,

$$\sigma = \varepsilon_o E_o (1 - 1/n^2) \sin\theta_B. \tag{17}$$

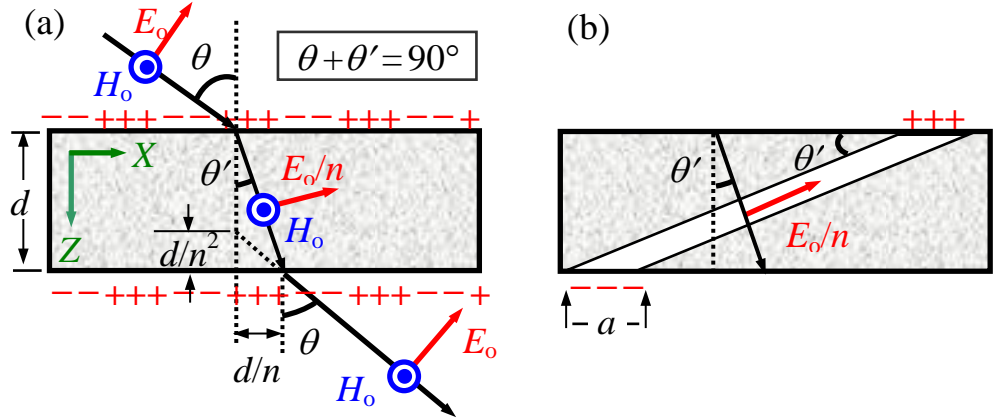

Fig. 6. (a) Dielectric slab of thickness $d$ and index $n$, illuminated with a *p*-polarized plane wave at Brewster's angle $\theta_B$. The *H*-field's magnitude inside the slab is the same as that outside, but the *E*-field inside is reduced by a factor of $n$ compared to the outside field. The bound charges on the upper and lower surfaces feel the force of the *E*-field. The transmitted beam is displaced by $d/n$ horizontally and by $d(1 - 1/n^2)$ vertically. The force $F_x\boldsymbol{x} + F_z\boldsymbol{z}$ on the upper surface is cancelled out by the force on the lower surface, but the slab experiences a net torque from $F_x$. The torque of $F_z$ is cancelled out by the forces exerted at the beam's edges within the dielectric. (b) Tilted cylinder of base area $a$ and length $d/\sin\theta'_B$, aligned with internal *E*-field.

We digress momentarily to look at the induced charges on the two facets of the slab, and verify their consistency with Maxwell's equations. The polarization density is $\boldsymbol{P} = \varepsilon_o(\varepsilon - 1)\boldsymbol{E}$, where $\boldsymbol{E}$ inside the medium has magnitude $E_o/n$. Consider a (tilted) cylindrical volume, aligned with the internal $E$-field and stretched between the two facets, as shown in Fig. 6(b). If the base area of this cylinder is denoted by $a$, its volume will be $ad$, where $d$ is the thickness of the slab. The electric dipole moment of this cylindrical volume is thus $ad\boldsymbol{P}$, which must be equal to the surface charge $a\sigma$ on either base multiplied by the length of the cylinder, $d/\sin\theta'_B = nd/\sin\theta_B$. The charge density on each surface is thus $\sigma = \varepsilon_o(\varepsilon - 1)E_o \sin\theta_B/n^2 = \varepsilon_o E_o(1 - 1/n^2)\sin\theta_B$, consistent with our earlier finding in Eq. (17).

In the horizontal direction, $E_x = E_o\cos\theta_B$ is continuous across the boundary, and the Lorentz force on the charges is

$$F_x = \tfrac{1}{2}\sigma E_x = \tfrac{1}{2}\varepsilon_o E_o^2 (1 - 1/n^2) \sin\theta_B \cos\theta_B. \tag{18}$$

The factor ½ accounts for time-averaging (also spatial-averaging over each surface, since the field and the charges vary sinusoidally both with time and with the coordinate $x$). If the incident beam's cross-sectional area is denoted by $A$, its footprint on the slab will have an area $A/\cos\theta_B$. The force density $F_x$, when integrated over the footprint and multiplied by the distance $d$ between the two surfaces, yields the following torque $T$ on the slab:

$$T = \tfrac{1}{2}\varepsilon_o E_o^2 Ad(1 - 1/n^2)\sin\theta_B. \tag{19}$$

Note that the net force on the slab is zero, because the top and bottom surfaces cancel each other out. However, the $F_x$ component yields a torque that tends to rotate the slab around the $y$-axis. Now, the electromagnetic field's momenta before and after the slab are the same, both in magnitude ($p = \tfrac{1}{2}A\varepsilon_o E_o^2$) and in direction, resulting in no net imparted force. However, upon transmission through the slab, the incoming $\boldsymbol{p}$ is displaced parallel to itself by $\Delta = d(1 - 1/n^2)\sin\theta_B$, as shown in Fig. 6(a). The change $p\Delta$ in the angular momentum of the beam is thus seen to be identical with the torque $T$ exerted by $F_x$, as given by Eq. (19). As a numerical example, consider the case of a glass slab having $A = 1.0$ mm$^2$, $d = 10$ μm, and $n = 1.5$, illuminated at $\theta_B = 56.3°$ with a 1.0 W/mm$^2$ beam of light. Using Eq. (19), the torque on the slab is found to be $T = 15.4$ f N.m.

Next, we consider the perpendicular force $F_z$ on the top surface of the slab. Since $E_z$ is discontinuous across the boundary, we must average $E_z$ just above and just below the interface. Thus $F_z = \tfrac{1}{2}\sigma E_z = \tfrac{1}{4}\varepsilon_o E_o^2(1 - 1/n^2)(1 + 1/n^2)\sin^2\theta_B$, with the factor ½ introduced again to account for time- and space-averaging. Integrating over the footprint of the beam (area = $A/\cos\theta_B$), we obtain $F_z = \tfrac{1}{4}\varepsilon_o E_o^2(n^2 - n^{-2})A\cos\theta_B$. The forces $F_z$ on the top and bottom facets of the slab, having equal magnitudes and opposite signs, cancel out.

Note that $F_z$, being laterally displaced by $d\tan\theta'_B = d/n$ between the top and bottom facets, must also exert a torque on the plate. This torque, however, is exactly cancelled out by an equal and opposite torque originating from the force of the beam exerted at its right and left edges. As derived in Section 6, the force density $F^{(edge)} = \tfrac{1}{4}\varepsilon_o(n^2 - 1)|E_t|^2$ is normal to the edge and expansive in the case of $p$-light. This force is of equal magnitude and opposite sign on the two edges of the beam inside the slab; while its horizontal components are aligned (and, therefore, do not give rise to a torque), its vertical components $F_z^{(edge)} = F^{(edge)}\sin\theta'_B$ produce a torque. The product of the area of each edge of the beam and the distance between the edges is $(d/\cos\theta'_B)(A/\cos\theta_B) = Ad(n^2 + 1)\cos\theta_B/\cos\theta'_B$, and the $E$-field inside the medium is $E_t = E_o/n$, so the torque produced by $F_z^{(edge)}$ is given by

$$T = \tfrac{1}{4}\varepsilon_o E_o^2 Ad\,(n - n^{-3})\cos\theta_B. \tag{20}$$

This is exactly equal and opposite to the torque exerted on the induced surface charges by the vertical component of the $E$-field.

**Note**: In a 1912 paper, G. Barlow (stimulated by J. H. Poynting) reports on an experiment similar to that described in this section, although incidence is not at Brewster's angle [16]. Barlow claims that his results confirm Poynting's previous theory, which apparently supports Minkowski's form of momentum inside the plate. His contention, however, cannot be correct as the torque is entirely determined by the momentum of the light outside the plate. Any momentum $p$ assigned to the light inside the glass medium results in the same overall torque.

### 8. Force experienced by an anti-reflection coating layer

To our knowledge, the discussion in this section has not appeared in the open literature. Aside from the practical significance of a force that can tear off a coating layer from its substrate, the results obtained below will be useful in balancing the electromagnetic and mechanical momenta of a beam of light that enters a dielectric slab from the free-space without any reflection losses at the interface.

A semi-infinite substrate of refractive index $n$ is shown in Fig. 7, coated with a dielectric layer of index $\sqrt{n}$ and thickness $d = \lambda_o/(4\sqrt{n})$. This quarter-wave-thick layer is a perfect anti-reflection coating that allows the incident beam to enter into the semi-infinite medium with no reflection whatsoever. Conservation of energy requires the *E*-field to enter the semi-infinite medium with an amplitude $|E_t| = E_o/\sqrt{n}$, while the quarter-wave thickness of the coating layer shifts the phase of $E_t$ by 90° relative to that of the incident beam. The *H*-field in the semi-infinite medium is then $H_t = nE_t/Z_o = i\sqrt{n}\, E_o/Z_o$. Inside the coating layer, the counter-propagating beams shown in Fig. 7 have the following distributions:

$$E_x(z) = \tfrac{1}{2}(1 + 1/\sqrt{n})\, E_o \exp(i2\pi\sqrt{n}\, z/\lambda_o) + \tfrac{1}{2}(1 - 1/\sqrt{n})\, E_o \exp(-i2\pi\sqrt{n}\, z/\lambda_o), \quad (21a)$$

$$H_y(z) = \tfrac{1}{2}(1 + \sqrt{n})\, H_o \exp(i2\pi\sqrt{n}\, z/\lambda_o) + \tfrac{1}{2}(1 - \sqrt{n})\, H_o \exp(-i2\pi\sqrt{n}\, z/\lambda_o). \quad (21b)$$

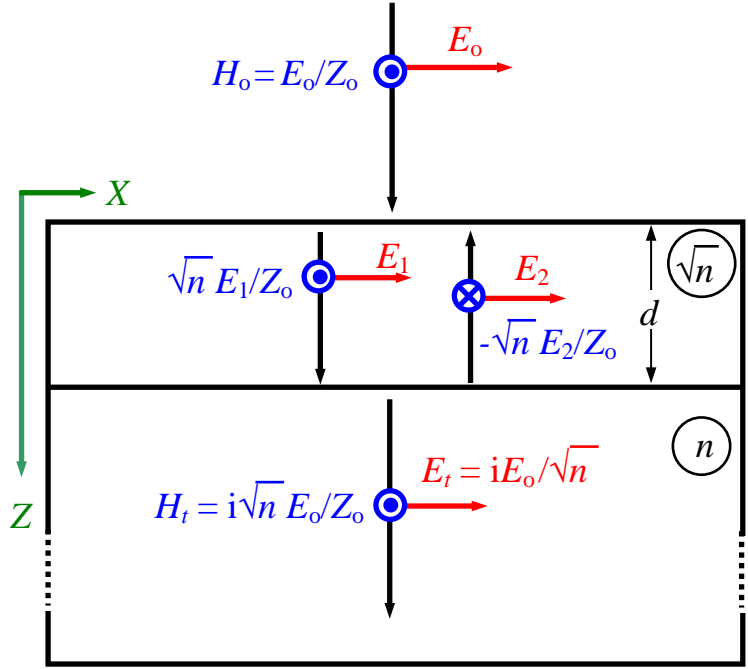

Fig. 7. Semi-infinite medium of index $n$, coated by a quarter-wave layer of index $\sqrt{n}$, a perfect anti-reflection layer. A normally incident plane wave is fully transmitted to the semi-infinite substrate. The standing wave within the coating layer produces an upward force equal to the time rate of change of the field's momentum in the free-space minus that in the substrate.

The force density, therefore, is given by

$$\begin{aligned} F_z &= \tfrac{1}{2}\, \text{Real}\, (J_x \times B_y^*) = \tfrac{1}{2}\, \text{Real}\, [-i\omega\varepsilon_o(\sqrt{\varepsilon} - 1)E_x \times \mu_o H_y^*] \\ &= [\pi(n-1)\sqrt{\mu_o\varepsilon_o}/\lambda_o]\, \text{Imag}\, (E_x \times H_y^*) \\ &= -\tfrac{1}{2}\varepsilon_o[\pi(n-1)^2/(\sqrt{n}\,\lambda_o)]\, \sin(4\pi\sqrt{n}\, z/\lambda_o)E_o^2. \end{aligned} \quad (22)$$

The above force density must be integrated over the thickness of the coating layer, from $z = 0$ to $z = \lambda_o/(4\sqrt{n})$, to yield the total force per unit area exerted on the layer, namely,

$$F_z = -\tfrac{1}{4}\, [(n-1)^2/n]\varepsilon_o E_o^2. \quad (23)$$

Considering that the incident momentum per unit time is $dp_i/dt = \tfrac{1}{2}\varepsilon_o E_o^2$, and that the time rate of change of the transmitted momentum into the semi-infinite medium is $dp_t/dt = \tfrac{1}{4}\varepsilon_o(\varepsilon + 1)|E_t|^2 = \tfrac{1}{4}\varepsilon_o(n^2 + 1)E_o^2/n$, it is clear that $F_z = d(p_i - p_t)/dt$; in other words, the upward force experienced by the coating layer is exactly equal to the time rate of change of the light's linear momentum upon crossing the layer. As a numerical example, consider the case of a glass substrate of index $n = 2.0$, coated with a quarter-wave thick layer of index $\sqrt{n} = 1.414$. At the incident power density of 1.0 W/mm$^2$, the computed net force on the coating layer is $F_z = -0.83$ nN/mm$^2$.

## 9. Homogeneous slab in free space

The role of interference fringes in creating a magnetic Lorentz force is discussed in the present section. As far as we know, this topic has not been covered in the open literature. Figure 8 shows a slab of thickness *d* and complex refractive index $n + i\kappa = \sqrt{\varepsilon}$, surrounded by free-space and illuminated at normal incidence. The slab's (complex) reflection and transmission coefficients are denoted by *r* and *t*, respectively. The counter-propagating beams within the slab have *E*-field amplitudes $E_1$ and $E_2$; the total field distribution is given by

$$E_x(z) = E_1 \exp(i2\pi\sqrt{\varepsilon}\, z/\lambda_o) + E_2 \exp(-i2\pi\sqrt{\varepsilon}\, z/\lambda_o) \quad (24a)$$

$$H_y(z) = (\sqrt{\varepsilon}\, E_1/Z_o) \exp(i2\pi\sqrt{\varepsilon}\, z/\lambda_o) - (\sqrt{\varepsilon}\, E_2/Z_o) \exp(-i2\pi\sqrt{\varepsilon}\, z/\lambda_o) \quad (24b)$$

Defining $\rho = [(\sqrt{\varepsilon} - 1)/(\sqrt{\varepsilon} + 1)] \exp(i4\pi\sqrt{\varepsilon}\, d/\lambda_o)$, the various parameters of the system of Fig. 8 may be written as follows:

$$E_1 = 2E_o / [(1 + \rho) + \sqrt{\varepsilon}\,(1 - \rho)] \quad (25a)$$

$$E_2 = 2\rho E_o / [(1 + \rho) + \sqrt{\varepsilon}\,(1 - \rho)] \quad (25b)$$

$$r = \{\rho - [(\sqrt{\varepsilon} - 1)/(\sqrt{\varepsilon} + 1)]\}/\{1 - \rho\,[(\sqrt{\varepsilon} - 1)/(\sqrt{\varepsilon} + 1)]\} \quad (25c)$$

$$t = 4\sqrt{\varepsilon} / [(\sqrt{\varepsilon} + 1)^2 \exp(-i2\pi\sqrt{\varepsilon}\, d/\lambda_o) - (\sqrt{\varepsilon} - 1)^2 \exp(+i2\pi\sqrt{\varepsilon}\, d/\lambda_o)] \quad (25d)$$

The force per unit area of the slab can be calculated in a way similar to that described in the preceding section, namely, by calculating the Lorentz force of the $H$-field on the dipolar current distribution, then integrating through the thickness of the slab. The final result is

$$F_z = \tfrac{1}{4} \varepsilon_o(n^2 + \kappa^2 + 1) \left[1 - \exp(-4\pi\kappa d/\lambda_o)\right] |E_1|^2$$
$$+ \tfrac{1}{4} \varepsilon_o(n^2 + \kappa^2 + 1) \left[1 - \exp(+4\pi\kappa d/\lambda_o)\right] |E_2|^2$$
$$+ \tfrac{1}{2} \varepsilon_o(n^2 + \kappa^2 - 1) \left[\cos(4\pi n d/\lambda_o + \phi_{E1} - \phi_{E2}) - \cos(\phi_{E1} - \phi_{E2})\right] |E_1 E_2| \quad (26)$$

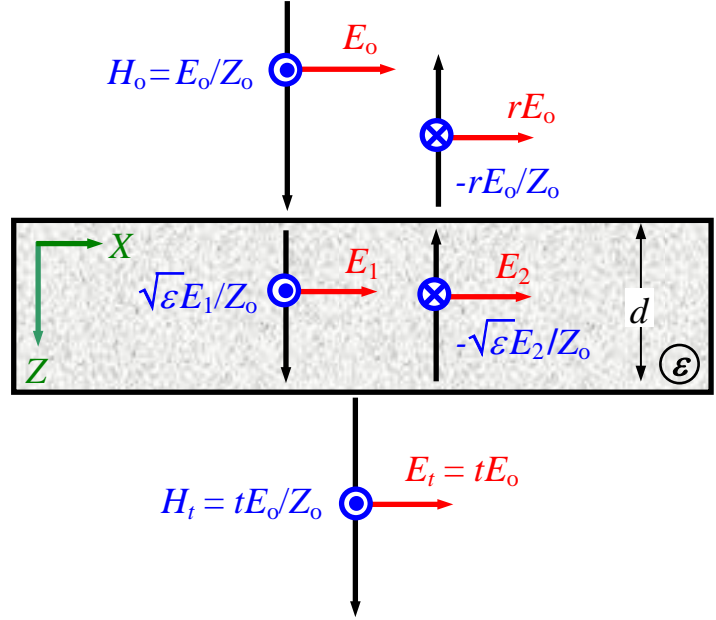

Fig. 8. A plate of thickness $d$ and dielectric constant $\varepsilon$ is illuminated at normal incidence by a linearly-polarized plane wave. The magnetic field's Lorentz force on the standing wave within the slab gives rise to a downward force that is precisely equal to the difference between the rates of incoming and outgoing momenta of the light beam.

In the special case when $\varepsilon$ is real, $\kappa$ goes to zero and the above formula reduces to

$$F_z = \frac{\varepsilon_o E_o^2}{1 + \left[\dfrac{2n}{(n^2 - 1)\sin(2\pi n d/\lambda_o)}\right]^2}. \quad (27)$$

This can be shown to be consistent with the balance of the incident, reflected, and transmitted momenta, namely,

$$F_z = \tfrac{1}{2} \varepsilon_o E_o^2 (1 + |r|^2 - |t|^2) = \varepsilon_o |r|^2 E_o^2, \quad (28)$$

where we have used $|r|^2 + |t|^2 = 1$ for a non-absorbing slab. According to Eq. (27), the net force is zero when $d = \lambda_o/2n$. This is consistent with the fact that a half-wave-thick plate does not reflect at all and, therefore, the incident and transmitted momenta are identical. The force density is a maximum for a quarter-wave-thick slab, and is given by

$$F_z = \varepsilon_o[(n^2 - 1)/(n^2 + 1)]^2 E_o^2. \tag{29}$$

At the incident power density of 1.0 W/mm$^2$, the radiation pressure on a quarter-wave slab of $n = 1.5$ is thus predicted to be 0.98 nN/mm$^2$.

**10. Gaussian beam in homogeneous, isotropic medium**

Loudon [17] has discussed the general case of Laguerre-Gaussian beams in great detail. The results of the present section, however, differ substantially from Loudon's findings; in particular, whereas Loudon finds the force experienced by the dielectric host of a Gaussian beam to be compressive irrespective of the beam's polarization state, we find compressive forces for *s*- and expansive forces for *p*-polarized light. These differences can be traced to the formula used for calculating the *E*-field component of the Lorentz force. Alternative formulations of the Lorentz force and the conditions under which each can be considered valid are discussed in Section 15.

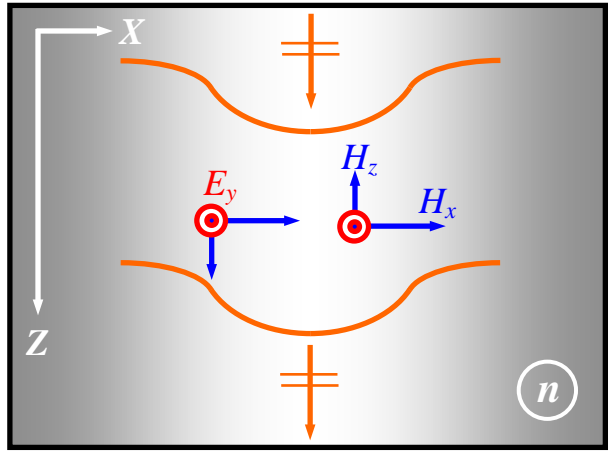

Fig. 9. One-dimensional Gaussian beam propagating along the *z*-axis in a dielectric medium of refractive index *n*. For the *s*-light shown here, the *E*-field has one component, $E_y$, while the *H*-field has both $H_x$ and $H_z$. The lateral force $F_x$ is positive on the left- and negative on the right-hand side, producing compressive pressure toward the beam center. For *p*-light (not shown), the field components are $H_y$, $E_x$, $E_z$, and the lateral force $F_x$, while having the same magnitude as in the case of *s*-light, is expansive in nature.

A one-dimensional Gaussian beam propagating along the *z*-axis in an isotropic, homogeneous medium, is shown in Fig. 9. The *E*-field in this system is along the *y*-axis, and has the following Gaussian profile:

$$E_y(z) = E_o\sqrt{\alpha(z)/\alpha(0)} \exp(i2\pi z/\lambda) \exp[-\alpha(z)x^2]. \tag{30}$$

Here $E_o$ is the magnitude of the field at the origin $(x, z) = (0, 0)$, $\lambda = \lambda_o/n$ is the wavelength inside the homogeneous dielectric of refractive index *n*, and $\alpha(z) = 1/r_o^2(z) - i\pi/\lambda R_c(z)$ is the Gaussian beam's complex parameter defined in terms of the beam's 1/e radius $r_o$ and radius of curvature $R_c$. The parameter $\alpha(z)$ evolves from its initial value $\alpha(0)$ according to the Gaussian beam formula $1/\alpha(z) = 1/\alpha(0) + i\lambda z/\pi$. The beam's *H*-field may be found from the Maxwell equation $\nabla \times \underline{E} = -\partial \underline{B}/\partial t$, or its time-harmonic equivalent $\nabla \times E = i\omega\mu_o H$. The only components of *H* turn out to be $H_x$ and $H_z$, given by

$$H_x = -(n/Z_o)E_o\sqrt{\alpha(z)/\alpha(0)}\{1 + (\lambda/2\pi)^2[2\alpha^2(z)x^2 - \alpha(z)]\}\exp(i2\pi z/\lambda)\exp[-\alpha(z)x^2], \tag{31a}$$

$$H_z = i(n/Z_o)E_o\sqrt{\alpha(z)/\alpha(0)}[(\lambda/\pi)\alpha(z)x]\exp(i2\pi z/\lambda)\exp[-\alpha(z)x^2]. \tag{31b}$$

The higher-order terms of $H_x$ diminish for a reasonably large beam size $r_o$, so only the first term in Eq. (31a) is significant. Also, since the Gaussian beam profile in Eq. (30) is valid only in the paraxial approximation, if the beam's *E*-field is re-calculated from the *H*-field components of Eq. (31) using $\nabla \times \boldsymbol{H} = -i\omega \varepsilon_o \varepsilon \boldsymbol{E}$, the $E_y(z)$ of Eq. (30) will be recovered only after higher-order terms similar to those in Eq. (31a) have been neglected.

Here we are concerned with the lateral force $F_x$ and, therefore, the only relevant fields are $E_y$ and $H_z$. The lateral component of force per unit volume along the *x*-axis is given by

$$F_x(z) = \tfrac{1}{2} \, Real(J_y B_z^*) = -\varepsilon_o E_o^2 (n^2 - 1) \, |\alpha(z)/\alpha(0)| \, [x/r_o^2(z)] \, \exp[-2x^2/r_o^2(z)]. \tag{32}$$

At the beam's waist $\alpha(z) = \alpha(0) = 1/r_o^2$, and Eq. (32) simplifies to

$$F_x(z = 0) = -\varepsilon_o E_o^2 (n^2 - 1) \, (x/r_o^2) \, \exp(-2x^2/r_o^2). \tag{33}$$

This force density being proportional to *x*, is positive on one side of the Gaussian beam and negative on the other side. It is also compressive, meaning that it tends to pull the dielectric toward the beam center at $x = 0$. If the force density of Eq. (33) is integrated from $x = 0$ to $\infty$, the magnitude of the net force on either side of the center turns out to be $\tfrac{1}{4}\varepsilon_o(n^2 - 1)E_o^2$, which is identical with the result found in Section 5 for an *s*-polarized beam.

A similar calculation may be performed for a *p*-polarized Gaussian beam, specified in terms of its only *H*-field component, namely,

$$H_y(z) = H_o\sqrt{\alpha(z)/\alpha(0)} \, \exp(i2\pi z/\lambda) \, \exp[-\alpha(z)x^2]. \tag{34}$$

The *E*-field components $E_x$ and $E_z$ are found (using $\nabla \times \boldsymbol{H} = -i\omega \varepsilon_o \varepsilon \boldsymbol{E}$) to be

$$E_x = (Z_o/n)H_o\sqrt{\alpha(z)/\alpha(0)} \, \{1 + (\lambda/2\pi)^2 \, [2\alpha^2(z)x^2 - \alpha(z)]\} \, \exp(i2\pi z/\lambda) \, \exp[-\alpha(z)x^2] \tag{35a}$$

$$E_z = -i \, (Z_o/n)H_o\sqrt{\alpha(z)/\alpha(0)} \, [(\lambda/\pi)\alpha(z)x] \, \exp(i2\pi z/\lambda) \, \exp[-\alpha(z)x^2]. \tag{35b}$$

Since we are concerned with the lateral force $F_x$, the relevant field components are $E_z$ and $H_y$. The lateral component of force per unit volume along the *x*-axis is then given by

$$F_x(z) = -\tfrac{1}{2} \, Real(J_z B_y^*) = \mu_o H_o^2 (1 - 1/n^2) \, |\alpha(z)/\alpha(0)| \, [x/r_o^2(z)] \, \exp[-2x^2/r_o^2(z)] \tag{36}$$

At the beam's waist $\alpha(z) = \alpha(0) = 1/r_o^2$, and the above formula simplifies to

$$F_x(z = 0) = \mu_o H_o^2 (1 - 1/n^2)(x/r_o^2) \, \exp(-2x^2/r_o^2). \tag{37}$$

This force density, being proportional to *x*, is positive on one side of the Gaussian beam and negative on the other side. It is also expansive, in the sense that it tends to push the dielectric away from the beam center at $x = 0$. If the force density of Eq. (37) is integrated from $x = 0$ to $\infty$, the magnitude of the net force on either side of the center turns out to be $\tfrac{1}{4}\mu_o(1 - 1/n^2)H_o^2$, which is identical to that found in Section 6 for a *p*-polarized plane wave, considering that, in the limit of large $r_o$, Eq. (35a) yields $E_x(x = 0, z = 0) = (Z_o/n)H_o$.

We have thus shown that the lateral pressure of a Gaussian beam on its host (dielectric) medium is compressive for *s*-light, expansive for *p*-light, and has the exact same magnitude of $\tfrac{1}{4}\varepsilon_o(n^2 - 1)E_o^2$ as was deduced based on momentum considerations in Sections 5 and 6.

## 11. Interference fringes and lateral radiation pressure

This section provides yet another demonstration of the expansive/contractive nature of the lateral pressure and its dependence on the state of polarization. It thus becomes clear that the results of the preceding sections are far from unique, and that lateral pressure can be readily produced and manipulated by the superposition of plane-waves of differing propagation directions and polarization states.

By now it must be amply clear that the force of light on a dielectric requires some sort of interference between two or more plane waves. Such interference generally produces a phase difference between the local *E*- and *H*-fields, thus resulting in a non-zero value of the magnetic Lorentz force. We have seen examples of this in Sections 8 and 9, where interference inside a dielectric layer of finite thickness gave rise to local forces. We now examine the case of two plane-waves, both polarized in the *p*-direction, that propagate at angles ±θ relative to the *z*-axis; see Fig.10. Assuming that the field amplitudes inside the medium of refractive index *n* are $E_o$ and $H_o = (n/Z_o)E_o$, the field distributions are written

$$E_x(x, z) = 2E_o\cos\theta \cos(2\pi n\, x\, \sin\theta/\lambda_o)\, \exp(i2\pi nz \cos\theta/\lambda_o), \tag{38a}$$

$$E_z(x, z) = -2iE_o\sin\theta \sin(2\pi nx\, \sin\theta/\lambda_o)\, \exp(i2\pi nz \cos\theta/\lambda_o), \tag{38b}$$

$$H_y(x, z) = 2H_o\cos(2\pi nx\, \sin\theta/\lambda_o)\, \exp(i2\pi nz \cos\theta/\lambda_o). \tag{38c}$$

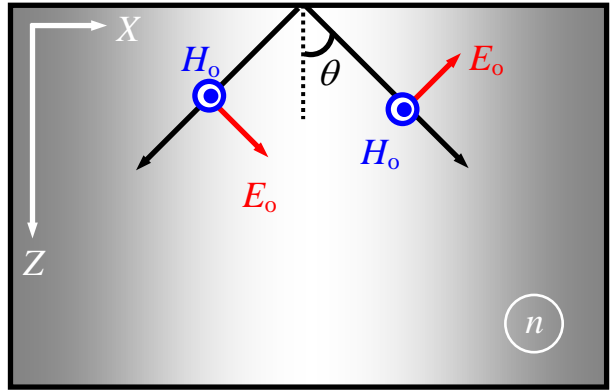

Fig. 10. Two linearly-polarized plane waves in a medium of refractive index *n* create interference fringes parallel to the *z*-axis. The lateral pressure on individual bright fringes is expansive in the case of *p*-polarization, and compressive in the case of *s*-polarization.

Since $E_x$ and $H_y$ are in-phase, they do not contribute to the force component along the *z*-axis; the only force component is

$$F_x = -\tfrac{1}{2}\, Real(J_z\, B_y^*) = (2\pi n\, \sin\theta/\lambda_o)\, (n^2 - 1)\varepsilon_o E_o^2\, \sin(4\pi nx\, \sin\theta/\lambda_o). \tag{39}$$

Within each fringe, this force is expansive, trying to stretch the dielectric away from the center of the bright fringe.

A similar analysis can be conducted for *s*-polarized plane waves, where the fields are

$$E_y(x, z) = 2E_o\cos(2\pi nx\, \sin\theta/\lambda_o)\, \exp(i2\pi nz \cos\theta/\lambda_o), \tag{40a}$$

$$H_x(x, z) = -2H_o\cos\theta\, \cos(2\pi nx\, \sin\theta/\lambda_o)\, \exp(i2\pi nz \cos\theta/\lambda_o), \tag{40b}$$

$$H_z(x, z) = 2iH_o\sin\theta\, \sin(2\pi nx\, \sin\theta/\lambda_o)\, \exp(i2\pi nz \cos\theta/\lambda_o). \tag{40c}$$

The only force component $F_x$ turns out to be

$$F_x = \tfrac{1}{2} Real(J_y\, B_z^*) = -(2\pi n\, \sin\theta/\lambda_o)\, (n^2 - 1)\varepsilon_o E_o^2\, \sin(4\pi nx\, \sin\theta/\lambda_o). \tag{41}$$

Within each fringe, this force is compressive, trying to pull the dielectric medium toward the center of the bright fringe. Aside from a minus sign, the forces in Eqs. (39) and (41) are identical. When this force is integrated over one-half of one fringe, say, from $x = 0$ to $x = \lambda_o/(4n\, \sin\theta)$, the result turns out to be

$$F_x = \pm\tfrac{1}{4}\, \varepsilon_o(n^2 - 1)(2E_o)^2. \tag{42}$$

The formula is written in the above form because the total *E*-field at the origin is $2E_o$. The force on each side of the fringe is thus seen to be identical to that for a Gaussian beam studied in Section 10, or for plane waves studied in Sections 5 and 6.

**12. Light pulse and the photon momentum**

In this section we clarify the notion of the mechanical momentum of light, the expression for which was derived in Section 4. At the same time, we make contact with the results of Loudon [6,7], who arrived at similar conclusions based on his quantum mechanical treatment of the problem. Although we agree with Loudon's expression for the electromagnetic momentum of a photon inside a dielectric, our mechanical momentum turns out to be somewhat greater than that indicated by his formula. We believe this disagreement is rooted in Loudon's neglect of the effect of an anti-reflection coating layer on the mechanical momentum imparted to the dielectric medium.

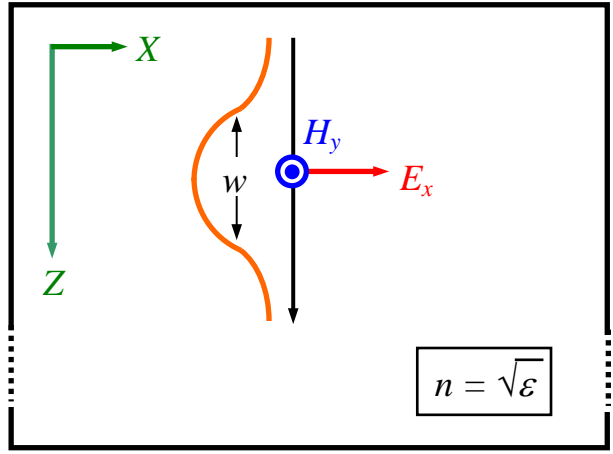

Fig. 11. A pulse of light (free-space wavelength = $\lambda_o$) travels along *z* in a dielectric of refractive index *n*. The light amplitude distribution in the *xy*-plane is uniform, similar to that of a plane wave, but the beam profile along *z* is Gaussian. The beam is linearly polarized, with its *E*-field along the *x*-axis and *H*-field along the *y*-axis. The assumed non-dispersive nature of the medium ensures that the pulse's group velocity is equal to the phase velocity *c/n*.

Figure 11 shows a light pulse in the form of a normally-incident plane wave entering a non-dispersive dielectric slab of refractive index *n*. For concreteness, we assume a Gaussian temporal profile for the pulse, although any other profile should, in the end, lead to the same conclusions. The *E*- and *H*-fields are thus given by

$$E_x(z, t) = E_o \exp\{-[(nz - ct)/w]^2\} \cos[2\pi(nz - ct)/\lambda_o], \tag{43a}$$

$$H_y(z, t) = (n/Z_o) E_x(z, t). \tag{43b}$$

Here $E_o$ is the amplitude of the *E*-field at the pulse center where $z = (c/n)t$, and *w* is a measure of the pulse width along the *z*-axis. The instantaneous density (per unit volume) of the Lorentz force, which has a component in the *z* direction only, is given by

$$F_z(z, t) = J_x(z, t) B_y(z, t) = \mu_o \varepsilon_o (\varepsilon - 1)(\partial E_x/\partial t) H_y(z, t)$$

$$= \varepsilon_o(\varepsilon - 1)E_o^2 \{2(n/w)[(nz - ct)/w]\cos^2[2\pi(nz - ct)/\lambda_o] + (\pi n/\lambda_o)\sin[4\pi(nz - ct)/\lambda_o]\}$$

$$\times \exp\{-2[(nz - ct)/w]^2\}. \tag{44}$$

Assuming $w \gg \lambda_o$, the local force density at each point $(x, y, z)$ can be time-averaged over one oscillation period. The squared cosine term in Eq. (44) thus averages to ½ and the sine term to zero, yielding the following expression for the average local force density:

$$\langle F_z(z, t)\rangle = \varepsilon_o(\varepsilon - 1)(n/w)[(n z - ct)/w]\, E_o^2 \exp\{-2[(n z - ct)/w]^2\}. \qquad (45)$$

The above force density can now be integrated on each side of the pulse center, say, from $z = (c/n)t$ to $\infty$, or from $z = -\infty$ to $(c/n)t$, to yield the total force (per unit cross-sectional area) at the leading and trailing edges of the pulse as follows:

$$F_z = \pm \tfrac{1}{4}\,\varepsilon_o(\varepsilon - 1)E_o^2. \qquad (46)$$

At the leading edge, where $F_z$ is positive, the pulse tends to push the medium forward, whereas at the trailing edge, where $F_z$ is negative, the medium is being pulled backward.

The Poynting vector $S$ is directed along the $z$-axis and, at the pulse center, has a magnitude $S_z = \tfrac{1}{2}(n/Z_o)E_o^2$. If the cross-sectional area of the beam is denoted by $A$ and the effective pulse duration by $\Delta T$, its total energy may be written as $S_z A \Delta T$. [The same result is obtained by adding the $E$-field and $H$-field energy densities, $\tfrac{1}{4}\varepsilon_o\varepsilon E_o^2$ and $\tfrac{1}{4}\mu_o H_o^2$, then multiplying with the effective pulse length $(c/n)\Delta T$ and also with the cross-sectional area $A$. Incidentally, for the Gaussian pulse of Eq. (43), $\Delta T = \sqrt{\pi/2}\, w/c$.] If the light pulse happens to represent a single photon, its total energy must be equal to $hf$, where $h$ is Planck's constant and $f$ is the optical frequency. The photon intensity is thus found to be $\tfrac{1}{2}E_o^2 = Z_o hf/(nA\Delta T)$.

When the pulse first enters the dielectric slab, the positive force of its leading edge accelerates the slab. The acceleration continues until the trailing edge enters, at which point the net force returns to zero. If the mass of the slab is denoted by $M$ – this could include the mass of the Earth, to which the slab is attached – its acquired momentum will be given by the integrated force over the pulse duration, namely, $MV = \tfrac{1}{4}\varepsilon_o(\varepsilon - 1)E_o^2 A \Delta T$. [This reduces to $MV = \tfrac{1}{2}(n^2 - 1)hf/nc$ for a single photon.] So long as the pulse stays within the slab this acquired momentum remains constant. However, as soon as the leading edge of the pulse exits through the slab's rear facet, the trailing edge begins to exert a braking force to slow down the slab's motion. By the time the trailing edge leaves the slab, the motion has come to a halt, and all the momentum initially acquired by the slab has returned to the light pulse.

In the last paragraph of Section 4 we argued that the correct expressions for the electromagnetic and mechanical momentum densities within a dielectric medium are $S/c^2$ and $\tfrac{1}{2}(\varepsilon - 1)S/c^2$, respectively. With the pulse fully contained within the slab, its effective length along the $z$-axis is $(c/n)\Delta T$ while its cross-sectional area is $A$; the total electromagnetic momentum of the pulse should, therefore, be $S_z A \Delta T/nc$. Since $S_z A \Delta T$ is the pulse's energy, which equals $hf$ for a single photon, the photon's electromagnetic momentum (when inside the slab) turns out to be $hf/nc$. By the same token, the photon's mechanical momentum (again, when fully contained within the slab) should be $\tfrac{1}{2}(n^2 - 1)hf/nc$. This is precisely the momentum $MV$ that the slab temporarily acquires from a single photon entry, as explained in the preceding paragraph. The physical basis for assigning a mechanical momentum to individual photons should thus be abundantly clear in light of the above considerations.

The sum of the electromagnetic and mechanical momenta of a photon inside a dielectric slab, $[(n^2 + 1)/(2n)]hf/c$, turns out to be greater than $hf/c$, the momentum of the same photon in free space. If the facets of the slab happen to be anti-reflection coated, the difference between the inside and outside momenta will be balanced by the force exerted by the entering and exiting photons on the coating layer (see Section 8 for a discussion of this force). In the absence of anti-reflection layers, the facets reflect a fraction of the photons, in which case the momentum transferred to the slab by the reflected photons can be used in a statistical analysis to account for the difference between the photon momenta inside and outside the slab.

## 13. Mirror immersed in liquid dielectric

Careful experiments [11] have shown that the ratio of the radiation pressure on a metallic reflector immersed in a variety of dielectric liquids to the radiation pressure on the same reflector in air is equal to the liquid's refractive index $n$. [The indices of the liquids used ranged from 1.33 (water) to 1.61 (carbon disulphide)]. In the present section we analyze the case of a perfect metallic conductor immersed in a liquid of arbitrary refractive index $n$, and show that a direct application of our method yields results that are in full agreement with the reported experiments.

With reference to Fig. 12, we define the reflection coefficient at the liquid surface (in the absence of the mirror) as $\rho = (1 - n)/(1 + n)$, and the single-path phase shift through the liquid column as $\phi = 2\pi n\, d/\lambda_o$. Upon matching the $E$- and $H$-fields at the liquid surface, we find

$$r = [\rho - \exp(i2\phi)] / [1 - \rho \exp(i2\phi)], \tag{47a}$$

$$E_t/E_o = (1 + \rho) / [1 - \rho \exp(i2\phi)]. \tag{47b}$$

It is easy to verify from Eq. (47a) that $|r| = 1$, as it should be, considering that neither the liquid nor the metallic reflector absorb the light. As for the radiation pressure on the mirror, since the magnitude of the surface current is $J_s = 2H_t = 2nE_t/Z_o$, the force per unit area at the mirror surface is given by

$$F_z^{(\text{Mirror})} = \tfrac{1}{2}\, \text{Real}\, (\boldsymbol{J}_s \times \mu_o \boldsymbol{H}_t^*) = [n^2/(\sin^2\phi + n^2\cos^2\phi)]\, \varepsilon_o E_o^2. \tag{48}$$

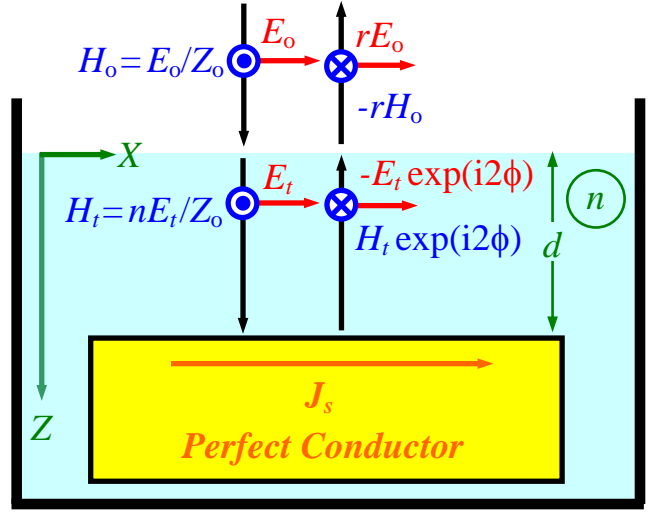

Fig. 12. A perfectly conducting mirror is immersed at depth $d$ below the surface in a liquid of refractive index $n$. The normally incident beam is a linearly-polarized plane wave of wavelength $\lambda_o$ and $E$-field magnitude $E_o$. The incident $E$-field's magnitude beneath the liquid surface is denoted by $E_t$. Upon reflection from the mirror the $E$-field retains its magnitude but undergoes a 180° phase shift. The reflected beam is also phase-shifted by $2\phi = 4\pi n\, d/\lambda_o$ in its round trip to the mirror and back. The induced surface current $\boldsymbol{J}_s$ at the top of the mirror is parallel to the $E$-field, but its magnitude is equal to the net $H$-field at the mirror surface. Interference between the incident and reflected beams creates standing-wave fringes within the liquid that exert a spatially varying, $\pm z$-directed force on the liquid. In the free-space region above the liquid, the reflected $E$- and $H$-fields have the same magnitudes as their incident counterparts, namely, $|r| = 1$. The relative phase between the incident and reflected fields (i.e., the phase of $r$) is determined by the boundary conditions at the liquid surface.

The radiation pressure on the immersed mirror is thus found to be greater than or equal to that experienced by the same mirror in free-space, namely, $\varepsilon_o E_o^2$. If the depth $d$ of the liquid column happens to be an integer-multiple of $\lambda_o/2n$, then $F_z$ in Eq. (48) turns out to be equal to the radiation pressure in free-space. However, if $d$ differs from the above value by $\lambda_o/4n$, the coefficient of $\varepsilon_o E_o^2$ in Eq. (48) acquires its maximum value of $n^2$. We conclude that, for an essentially monochromatic beam of light, the ratio of the radiation pressure on the immersed mirror to that in free-space could be anywhere between unity and $n^2$. In practice, the light source is never perfectly monochromatic and, in fact, its coherence length in many experiments is substantially smaller than the depth $d$ of the liquid. Under such circumstances, the radiation pressure is estimated by averaging $F_z$ of Eq. (48) over the source's bandwidth. Assuming equal likelihood for all values of $\phi$, the averaged coefficient of $\varepsilon_o E_o^2$ in Eq. (48) turns out to be equal to $n$, which is precisely what has been measured in the experiments.

Concerning the imbalance between the free-space momenta of the incident and reflected beams and the force experienced by the mirror, one must take into account the radiation pressure felt by the liquid in its interaction with the standing wave fringes between the mirror and the surface of the liquid. The $E$- and $H$-fields inside the liquid may be written similarly to those in the dielectric slab of Section 9 (see Eqs. (24)), and the magnetic Lorentz force on the bound currents can be straightforwardly calculated as follows:

$$F_z(z) = (2\pi n/\lambda_o)[(n^2 - 1)/(\sin^2\phi + n^2\cos^2\phi)] \, \varepsilon_o E_o^2 \sin[4\pi n(z-d)/\lambda_o]. \qquad (49)$$

The above (volume) force density, which varies sinusoidally between positive and negative values, averages out to zero within each fringe of the standing wave. When integrated from $z = 0$ to $d$, it yields the force per unit cross-sectional area of the liquid column as follows:

$$F_z^{(\text{Liquid})} = -[(n^2 - 1)\sin^2\phi/(\sin^2\phi + n^2\cos^2\phi)] \, \varepsilon_o E_o^2. \qquad (50)$$

Adding $F_z^{(\text{Mirror})}$ in Eq. (48) to $F_z^{(\text{Liquid})}$ in Eq. (50) reveals the net radiation pressure on the system of Fig. 12 to be $\varepsilon_o E_o^2$, in agreement with the time rate of change of the free-space momentum of the incident beam upon reflection from the liquid surface.

When using a light source with a coherence length shorter than the depth $d$ of the liquid column, the fringes beneath the liquid surface are washed out, but those closer to the mirror survive. These surviving fringes impart a net negative pressure to the liquid column that accounts for the aforementioned factor of $n$ increase in the radiation pressure experienced by the mirror. It is remarkable that the hydrostatic response of the liquid volume to this (nonuniform) pattern of radiation pressure does not appear to have affected the final result of the experiments.

**14. Optical tweezers**

It is a well-known experimental fact that a small glass bead is attracted to the center of a focused beam, irrespective of the beam's polarization state. The discussion in the preceding sections might lead one to conclude that while an $s$-polarized beam, by virtue of its compressive pressure, can pull a glass bead (laterally) to the point of highest intensity, a $p$-polarized beam, because of the expansive nature of its force, tends to push the glass bead away. In order to clarify this misunderstanding (in the case of $p$-light) let us examine the passage of a localized beam through a glass slab in the vicinity of one of the slab's sidewalls; see Fig. 13. Unlike the analyses elsewhere in the paper, the present discussion cannot be made rigorous without extensive numerical calculations, so we limit our analysis to a few qualitative remarks.

Suppose, as shown in Fig.13, that the beam passes through the slab without producing any additional reflected, refracted, or diffracted beams, which would interfere with the main beam and complicate the analysis. Let the $x$-component of the $E$-field in the vicinity of the sidewall in Fig. 13 have magnitude $E_o$ just outside the slab. The continuity of $D_\perp$ at this sidewall then requires that $E_x$ just inside the slab have a magnitude equal to $E_o/n^2$. The p-polarized beam inside the slab thus exerts an expansive force on the bulk in the positive x-direction, with a force per unit area $F^{(edge)} = \frac{1}{4}\varepsilon_o(n^2 - 1)(E_o/n^2)^2$. The (bound) charge density at the sidewall is obtained from the discontinuity in $E_\perp$ as $\sigma = -\varepsilon_o(1 - 1/n^2)E_o$. When the (average) E-field at the sidewall, $E^{(eff)} = \frac{1}{2}(1 + 1/n^2)E_o$, interacts with the above charge density, it gives rise to a force per unit area $F^{(wall)} = \frac{1}{2} Real\, (\sigma E^{(eff)*}) = -\frac{1}{4}\varepsilon_o(1 - 1/n^4)E_o^2$; the minus sign in front indicates that the force is along the negative x-axis. The net horizontal force, $F_x = F^{(wall)} + F^{(edge)} = -\frac{1}{4}\varepsilon_o(1 - 1/n^2)E_o^2$, thus pulls the slab toward the center of the beam, in agreement with the experimental observations.

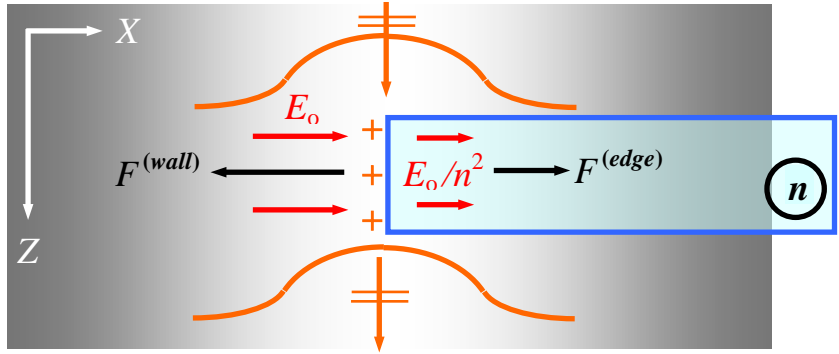

Fig. 13. A one-dimensional Gaussian beam, linearly-polarized along the x-axis, passes through a glass plate near one of the plate's sidewalls. The (bound) charges induced on the plate's wall experience a Lorentz force $F^{(wall)}$ from the E-field of the beam. This force is opposite in direction to the internal force $F^{(edge)}$ exerted on the bulk of the medium at the beam's edge that is inside the plate. Since $F^{(wall)}$ is generally stronger than $F^{(edge)}$, the net force tends to pull the plate to the left.

### 15. Concluding remarks

An important result of this paper is the expression for the momentum density $p$ of electromagnetic plane-waves in isotropic and homogeneous dielectric media. The old debate as to whether $p = \frac{1}{2}D \times B$ or $p = \frac{1}{2}E \times H/c^2$ has been settled by showing that the correct expression is obtained by averaging the two forms. We examined several cases involving a single plane-wave in a dielectric medium, and verified that the new expression for $p$ is valid in each case. The utility of the formula is limited, however, as in most cases of practical interest multiple plane-waves interfere with each other, and the force experienced by the media must then be calculated from a direct consideration of the (conduction and/or bound) charges and currents, and their interactions with the electric and magnetic components of the radiation field. Examples of these calculations were given, and the results in all cases were found to be consistent with the known properties of electromagnetic radiation. The case of a metallic mirror immersed in a dielectric liquid, which yields to exact analysis, was shown to be in agreement with reported experiments as well.

Another major result of this paper is the prediction that, inside dielectric media, an expansive lateral pressure exists at the edges of p-polarized, finite-diameter beams. We arrived at this result from four different perspectives in Sections 6, 7, 10, and 11, and showed that the magnitude, direction, and polarization-dependence of the lateral pressure are exactly the same in all the cases considered. In light of these analyses, it is plausible to argue that the strength of the lateral pressure is independent of the cross-sectional profile of the beam, in general, and of the detailed structure of the beam's edge, in particular.

Our method of calculating the radiation pressure can be used in conjunction with numerical simulations to yield the distribution of fields and forces in diverse systems of practical interest. We will report our numerical computation results in a forthcoming paper.

We close by pointing out that in some of the published literature [5,14,17] the point of departure for calculating the radiation pressure is the ponderomotive force density,

$$\boldsymbol{F} = (\boldsymbol{P}\cdot\nabla)\boldsymbol{E} + (d\boldsymbol{P}/dt)\times\boldsymbol{B}. \tag{51}$$

This is the Lorentz force of the radiation's $E$- and $B$-fields on the induced electric dipole density $\boldsymbol{P}$, which, in turn, is assumed to be proportional to the applied $E$-field, that is, $\boldsymbol{P} = \varepsilon_o(\varepsilon - 1)\boldsymbol{E}$. It is apparent that wherever both $\boldsymbol{P}$ and the $E$-field gradient happen to have non-zero projections along any coordinate axis, the medium experiences an electric Lorentz force. This, however, is *not* the same as the force of the *total* $E$-field on the induced charges, which is what we have been using throughout the paper. Whereas $\boldsymbol{E}$ in Eq. (51) is produced by external sources alone, our total $E$-field has included the contributions of the induced charges and currents within the dielectric medium as well. In the latter case, the Maxwell equation $\nabla\cdot\boldsymbol{D} = 0$ in conjunction with the constitutive relation $\boldsymbol{D} = \varepsilon_o\varepsilon\boldsymbol{E}$ ensures the absence of (unbalanced) bound charges from the interior of the medium; the only charges then are those induced at the interface(s) between adjacent media of differing dielectric constants. The difference between some of the results in this paper and those in the literature (e.g., the existence of an expansive lateral pressure at the edge of a beam) may thus be traced back to the *internal E*-field contribution to the Lorentz force, which is absent from Eq. (51), but has been properly accounted for in our formulation.

The rationale for expressing the $E$-field component of the Lorentz force in Eq. (51) as $(\boldsymbol{P}\cdot\nabla)\boldsymbol{E}$ may be understood if one considers the $E$-field's force on individual atoms, molecules, and even larger particles which, nonetheless, are small compared to the light's wavelength. In dealing with a dense continuum such as a solid or a liquid, however, one must avoid the use of ad hoc formulas and, instead, embrace the universal form of the Lorentz force, $\rho(\boldsymbol{r}, t)\boldsymbol{E}(\boldsymbol{r}, t)$, where $\rho$ is the local charge density. This approach assumes that the constituent atoms of the medium under consideration are vanishingly small, and that these atoms are uniformly and densely distributed throughout the volume. If one opts for the "clumpy" model of the material, in which individual atoms (or groups of atoms) occupy certain isolated locations in space, then the local $E$-field acting on each such clump will have to be calculated for the "clumpy situation" as well. This, however, will introduce variations in the $E$-field over and above the smooth distribution calculated from the macroscopic Maxwell equations. We thus believe that, if one uses the macroscopic equations to derive the optical $E$-field inside the medium, one must also ignore the "lumpiness" of the actual material and accept the smooth distribution of (bound) charges and dipole moments. This is the approach taken throughout this paper, in stark contrast to a majority of the published literature in the field, and is the source of the disagreement between some of our results and those published by others. While the published results obtained on the basis of Eq. (51) should remain applicable to low-density gases and dilute collections of small particles, it is our belief that, for dense, homogeneous solids and liquids, the approach taken in this paper is more suitable.

**Acknowledgments**

The author is grateful to Rodney Loudon, Ewan Wright, Pierre Meystre, Pavel Polynkin, and Kishan Dholakia for illuminating discussions. Professor Loudon also brought to my attention Planck's book and Barlow's paper, noted at the end of Sections 3 and 7. This work has been supported by the *Office of Naval Research* MURI grant No. N00014-03-1-0793, by the *National Science Foundation* STC Program under agreement DMR-0120967, and by the AFOSR contract F49620-02-1-0380 with the Joint Technology Office.